\documentclass[twocolumn]{aastex62}
\usepackage{amsmath}
\usepackage{hyperref}

\graphicspath{{./}{figures_pdf/}}

\received{November 27, 2018}
\revised{February 25, 2019}
\accepted{March 13, 2019}
\submitjournal{ApJ}

\shorttitle{AGN in BPT-SF Galaxies}
\shortauthors{Agostino and Salim}

\begin{document}
\title{Crossing the Line: Active Galactic Nuclei in the Star-forming region of the BPT Diagram}

\author{Christopher J. Agostino}
\affil{Indiana University\\
727 East 3rd St. Swain West 318\\
Bloomington, IN 47405-7105, USA}

\author{Samir Salim} 
\affil{Indiana University\\
727 East 3rd St. Swain West 318\\
Bloomington, IN 47405-7105, USA}
 
\begin{abstract}
  In this work, we investigate the reliability of the BPT diagram for excluding galaxies that host an AGN. We determine the prevalence of X-ray AGN in the star-forming region of the BPT diagram and discuss the reasons behind this apparent misclassification, focusing primarily on relatively massive ($\log(M_{*})\gtrsim10$) galaxies. X-ray AGN are selected from deep XMM observations using a new method that results in greater samples with a wider range of X-ray luminosities, complete to $\log(L_{X})>41$ for $z<0.3$. Taking X-ray detectability into account, we find the average fraction of X-ray AGN in the BPT star-forming branch is 2\%, suggesting the BPT diagram can provide a reasonably clean sample of star-forming galaxies. However, the X-ray selection is itself rather incomplete. At the tip of the AGN branch of the BPT diagram, the X-ray AGN fraction is only $14\%$, which may have implications for studies that exclude AGN based only on X-ray observations. Interestingly, the X-ray AGN fractions are similar for Seyfert and LINER populations, consistent with LINERs being true AGN. We find that neither the star-formation dilution nor the hidden broad-line components can satisfactorily explain the apparent misclassification of X-ray AGN. On the other hand, $\sim40\%$ of all X-ray AGN have weak emission lines such that they cannot be placed on the BPT diagram at all and often have low specific SFRs. Therefore, the most likely explanation for ``misclassified'' X-ray AGN is that they have intrinsically weak AGN lines, and are only placeable on the BPT diagram when they tend to have high specific SFRs.
  
 \end{abstract}


\section{Introduction} \label{sec:intro}

 The ability to reliably select or eliminate AGN in a sample of galaxies is often necessary for studying the evolution of galaxies and the study of the AGN themselves.  An incomplete selection of AGN could preclude us from understanding all aspects of the AGN phenomenon, whereas incomplete elimination of AGN could contaminate the measurements of host properties, such as the star formation rate (SFR) or gas-phase metallicity and would make the studies that contrast non-AGN with AGN less reliable.
  
 Under the Unified model of AGN, the orientation of an AGN with respect to the dust-obscuring torus in its host galaxy affects how an AGN appears to an observer \citep{antonucci1993}. When viewed directly, narrow forbidden lines and broadened Balmer lines are detected in the optical and UV bands, indicative of gas speeds in the broad-line region (BLR) on the order of thousands of kilometers per second. Such AGN are referred to as type 1. When viewed at an angle where light from the BLR is largely attenuated by some obscuring torus, narrow forbidden emission and Balmer recombination lines are seen from less dense gas that is further out from the central accretion disk of the AGN. Such AGN with strong forbidden emission lines compared to Balmer recombination lines are referred to as type 2 AGN. This simple picture has subsequently been expanded to recognize the AGN luminosity as an additional determining factor. Namely, lower-luminosity AGN will not be able to form a BLR, and would only feature narrow lines \citep{laor2003}. Such objects are sometimes referred to as true type 2 AGN and may constitute 7\% of type 2 AGN \citep{pons2016b}. Altogether, type 2 AGN are many times more common than type 1 AGN.

 In addition to the orientation, the detection, and consequently, the selection of AGN depends on their intrinsic properties, such as its Eddington ratio and the AGN luminosity \citep{hickox2009}. Type 1 AGN can be identified based on their extreme Balmer line broadening or their blue continuum. Type 2 AGN will neither feature broad Balmer lines nor will have a continuum different from non-AGN galaxies, so a variety of methods have been developed in an attempt to identify them. The Baldwin-Phillips-Terlevich \citep[BPT,][]{bpt1981} diagram is of particular interest because it uses commonly available optical emission line ratios, [OIII]$\lambda$5007/H$\beta$ and [NII]$\lambda$6583/H$\alpha$, to determine the source of ionization in nebular gas. The BPT diagram's two main features, the star-forming and AGN branches, are a result of gas-phase abundance, physical conditions of the ISM, and the processes that drive the ionization of nebular regions in galaxies. Galaxies in the star-forming branch have emission line ratios resembling those of individual HII regions, a result reproduced by photoionization models where O and B stars are the sources of ionization \citep{kewley2001, nagao2006, stasinska2006, levesque2010, sanchez2015}. Galaxies along the AGN branch have higher ratios of [OIII]/H$\beta$ and [NII]/H$\alpha$ because the higher energy photons produced in accretion processes can induce more heating in the narrow-line region (NLR), causing collisionally excited lines to emit more strongly relative to the Balmer recombination lines \citep{stasinska2006}. The AGN branch stems from the bottom of the star-forming branch to the upper right, because luminous AGN are primarily found in massive, metal-rich galaxies \citep{kauffmann2003, schawinski2010, mullaney2012}.
 
 \citet{kauffmann2003} used the BPT diagnostic to empirically derive a boundary to distinguish AGN from star-forming galaxies. Because [NII] and H$\alpha$ become difficult to measure at higher redshifts, other optical emission-line diagrams have been proposed as alternatives to the BPT diagnostic by swapping the [NII]/H$\alpha$ line-ratio for other proxies. \citet{vo87} used [SII]$\lambda \lambda$6717,6731/H$\alpha$ or [OI]$\lambda$6300/H$\alpha$. Other methods utilized host properties like stellar mass \citep{juneau2011} or rest $U-B$ color \citep{yan2011}. \citet{tbt2011} (TBT) used rest $g-z$ color and [NeIII]$\lambda$3869/[OII]$\lambda \lambda3826,3729$ to separate AGN from star-forming galaxies. Nevertheless, the BPT diagram is by far the most widely used emission-line diagram and may be the most reliable \citep{stasinska2006}.

 The reliability and the completeness of BPT classification have not been robustly explored. There are several manifestations of reliability. One is whether the galaxies hosting the AGN are identified as such. For example, \citet{trump2015} modeled a sample of AGN with a distribution of physically motivated Eddington ratios to show it is possible that the BPT diagram based on SDSS fiber spectroscopy is biased against AGN in low-mass, blue galaxies with high specific star-formation rates (sSFRs, star-formation rates normalized by the stellar mass of galaxies). By simulating a set of AGN with Eddington ratios distributed as a Schechter function, \citet{jones2016} similarly show that narrow-line AGN may be overpowered by host star formation, biasing the BPT classification of weaker AGN. 
 
 Another crucial aspect of reliability is whether the galaxies classified with the BPT diagram as AGN are true AGN. For example, it has been proposed that most galaxies exhibiting low-ionization nuclear emission-line regions \citep[LINERs,][]{heckman1980} are not ionized primarily by AGN \citep{stasinska2008, yan2012, belfiore2016}. The current paper will mostly focus on the first question, and will touch upon the second.

 Other approaches were developed that identify AGN (of either type) by excess emission in the radio \citep{condon1992}, IR \citep{kennicutt1998}, or X-ray \citep{ranalli2003}, allowing one to test the reliability of the BPT diagram. The majority of AGN are actually radio-quiet \citep{wilson1995} and separating such weak radio cores from host star-formation typically requires deep interferometric imaging with arcsecond-scale angular resolution \citep{ho2008}, rendering large statistical studies based on radio emission observationally expensive. IR selection often utilizes color cuts based on mid-IR colors to select AGN \citep{lacy2004, stern2005, mateos2012, stern2012}, but these selection methods can be affected by dust heating from old stars. By contrast, X-ray sources are typically selected using a luminosity cut of $\log(L_{\mathrm{X}})>42$, because any source more luminous than that limit is more likely to be an AGN than a star-forming galaxy \citep{fabbiano1989, zezas1998, moran1999} since the star-formation rates necessary to reach such high X-ray luminosities are $\sim 200 M_{\odot}$ yr$^{-1}$ \citep{ranalli2003}. The spatial density of such starbursting galaxies is quite low in the local ($z<0.3$) universe, leaving the X-ray method as the preferred one for testing the placement of secure AGN on the BPT diagram. A number of studies performed at low redshift have found that X-ray selected AGN, in addition to mostly populating the AGN branch of the BPT diagram, occasionally cross into the star-forming region of the BPT diagram \citep{trouille2010, tbt2011, castello2012, pons2014, pons2016a, pons2016b}, suggesting that the BPT classification is to some degree unreliable, leading to incomplete AGN selection, or alternatively, to contamination of ``non-AGN'' samples. 
 
 Various explanations have been proposed to account for the apparent misclassification by the BPT diagram. \citet{castello2012} used X-ray spectral analysis to argue that the majority of BPT star-forming X-ray AGN are actually narrow-line Seyfert 1s (NLS1), a subclass of type 1 AGN \citep{osterbrock1985} which exhibits narrow-broad composite lines, and whose broad component may distort the line ratios and lead to misclassification. In contrast, other studies have found NLS1s to be only a fraction of misclassified X-ray AGN \citep{pons2014, pons2016b}. \citet{pons2014} suggested that non-NLS1 X-ray AGN are misclassified because they either have significant contributions from star-formation by their host and/or low accretion rates.
 
 Previous studies have not reached a consensus on the reasons for misclassification, prompting us to revisit the question with more complete samples, more extensive ancillary data, and some new analysis approaches. Specifically, we use a novel selection method that produces a more complete sample of X-ray AGN at lower luminosities, and couple it with more accurate characterization of host galaxies in terms of their SFRs and stellar masses. Following previous studies, we focus on low redshifts ($z < 0.3$), where the quality of the data is high and samples are large. We primarily explore causes of misclassification for relatively massive ($\log(M_{*})\gtrsim 10$) galaxies, which contain most massive SMBHs and whose role in galaxy evolution may be more critical. On the other hand, the misclassification of AGN in low-mass, metal-poor galaxies may have fundamentally different causes and we are not discussing it in this work \citep{cann2019, dickey2019}.
 
 In addition to investigating the reasons for apparent misclassification, our goal is also to provide a statistically robust assessment of the contamination rate of AGN in the star-forming region of the BPT diagram. To assess the contamination, we calculate the fraction of X-ray AGN across the entire BPT diagram, taking into account their detectability as X-ray sources given the depth of the observations. This exercise also allows us to weigh in on the question of whether LINERs are true AGN.

\section{Sample and Data}  \label{sec:sample}
 In order to study the optical emission line properties of X-ray selected AGN, we need extensive but deep X-ray coverage to find a statistical sample of X-ray AGN candidates, accurate star formation rates for determining if the source of X-ray emission is indeed an X-ray AGN, and carefully measured optical emission-line fluxes to construct emission-line diagrams.
\subsection{Data}\label{sec:data}
 X-ray sources used in this work were compiled in the sixth data release of the third XMM-Newton serendipitous source catalog \citep[3XMM,][]{rosen2016}. 3XMM consists of nearly half a million sources identified in publicly available XMM-Newton observations. X-ray fluxes in 3XMM are given in several bands covering the energy interval of 0.2-12 keV, assuming a power-law spectral model with photon index $\Gamma=1.7$. In order to utilize the empirical relationships between star-formation-rate and X-ray luminosity for the soft (0.5-2 keV) and hard (2-10 keV) X-ray bands determined in \citet{ranalli2003}, fluxes in 0.5-10 keV band are obtained from 0.5-12 keV fluxes by multiplying by 0.9, a factor based on $\Gamma=1.7$, and are then converted into luminosities. 
 
  Star formation rates used in this work originate from GALEX-SDSS-WISE Legacy Catalog \citep[GSWLC,][]{salim2016}\footnote{\url{http://pages.iu.edu/~salims/gswlc/}}, an optically selected catalog of $\sim700,000$ galaxies with 100 Myr-averaged SFRs determined via UV/optical SED fitting. GSWLC is particularly useful in that it contains relatively robust SFRs even for galaxies off the galaxy star-formation (SF) main sequence, where many AGN are found. We utilize the medium depth catalog available from GSWLC, which we will hereby refer to as GSWLC-M1. GSWLC-M1, which covers approximately 4500 sq. deg., has a greater overlap with 3XMM catalog than the deeper but less extensive GSWLC-D1 catalog. Sources in 3XMM were matched to galaxies in GSWLC-M1 with a 7$\arcsec$ search radius. For cases where an X-ray source was within 7$\arcsec$ of multiple GSWLC-M1 sources, the GSWLC-M1 source with the highest $r$-band flux was adopted as a match. We found that 2041 3XMM sources have counterparts in GSWLC-M1.
  
\begin{figure}[t!]
                    \epsscale{1.2}
                    \plotone{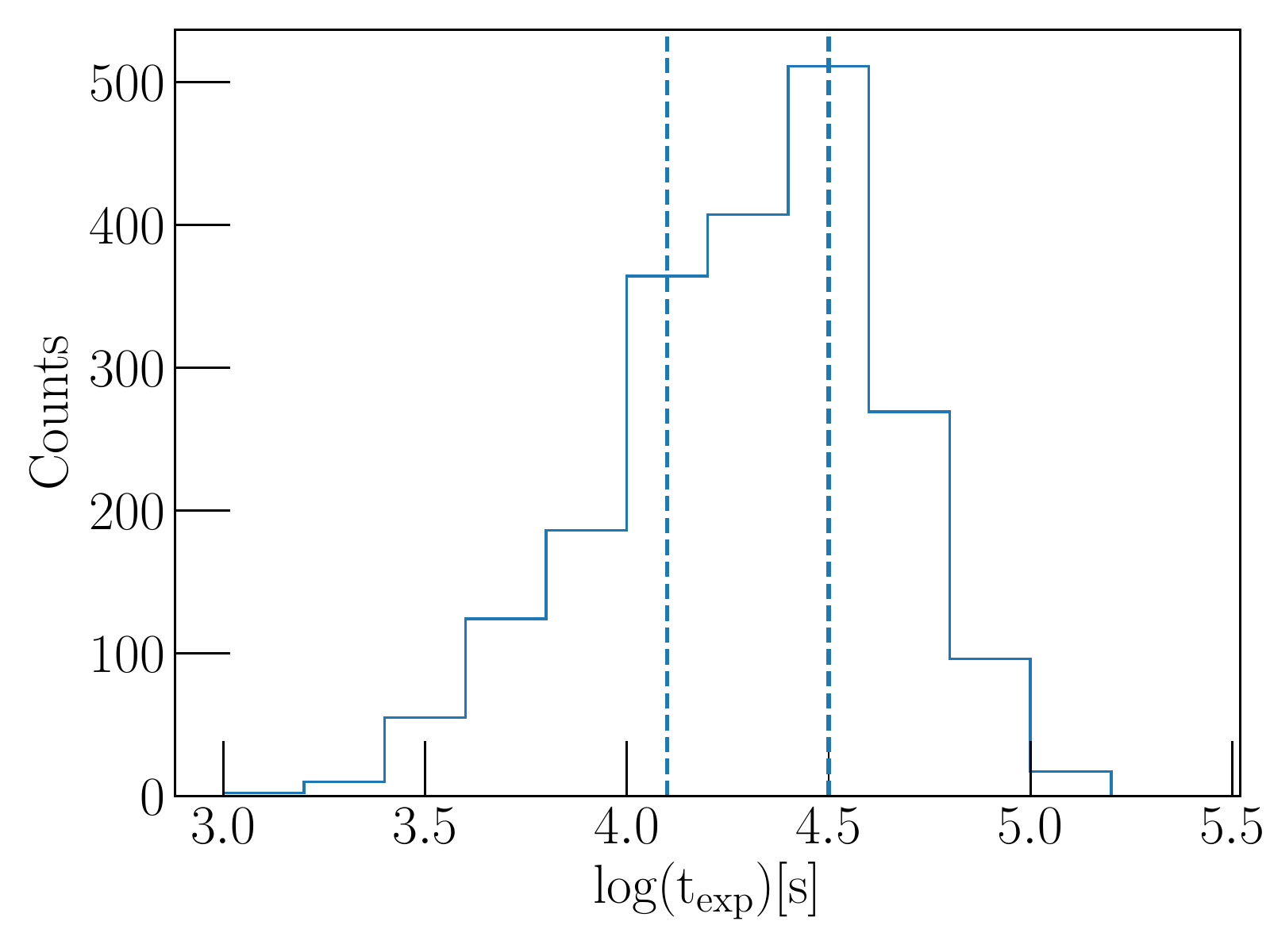}
                    \caption{Histogram of exposure times of observations for galaxies detected in 3XMM with matched optical counterparts in GSWLC-M1. Dashed lines denote the lower and upper limits of observation times used in our sample. \label{fig:exptimes}}
                    \end{figure}

 In order to have a consistent detection limit, we only use X-ray sources which were observed with exposure times falling within a relatively small range. In Figure \ref{fig:exptimes}, we show a histogram of the logarithm of exposure times of the 2041 3XMM sources with optical counterparts in GSWLC-M1. We retained 837 sources, which had $4.1<\log(t_{\mathrm{exp}})<4.5$, corresponding to $\sim20$ ks. We then required that each galaxy has a good quality SED fit, as defined by GSWLC-1, and we also removed type 1 AGN, i.e. galaxies classified spectroscopically in SDSS pipeline as `QSOs'. The final X-ray sample contained 740 galaxies.

 Spectral line fluxes used in this work originate from the MPA/JHU catalog of emission line measurements\footnote{\url{http://www.mpa-garching.mpg.de/SDSS/DR7/}}, derived from the spectra released in SDSS DR7 \citep{abazajian2009}  following \citet{tremonti2004}, and were available for 681 of the 740 galaxies due to the smaller coverage of DR7 with respect to DR10 used for the construction of GSWLC. Line fluxes were dust-corrected using the Balmer decrement and the Cardelli dust attenuation law \citep{cardelli1989}. 
 
 X-ray luminosities in our sample extend down to $\log(L_{\mathrm{X}})\sim 40$ for nearby galaxies and are complete for $\log(L_{\mathrm{X}})>41$ at all redshifts. The galaxies in the final sample span the redshift range defined by GSWLC, $0.01<z<0.3$ and the stellar masses are typically in $10<\log(M_{*})<12$ range.

\subsection{Selection of X-ray AGN sample}\label{sec:xrsfr}

\begin{figure}[t!]
                    \epsscale{1.3}
                    \plotone{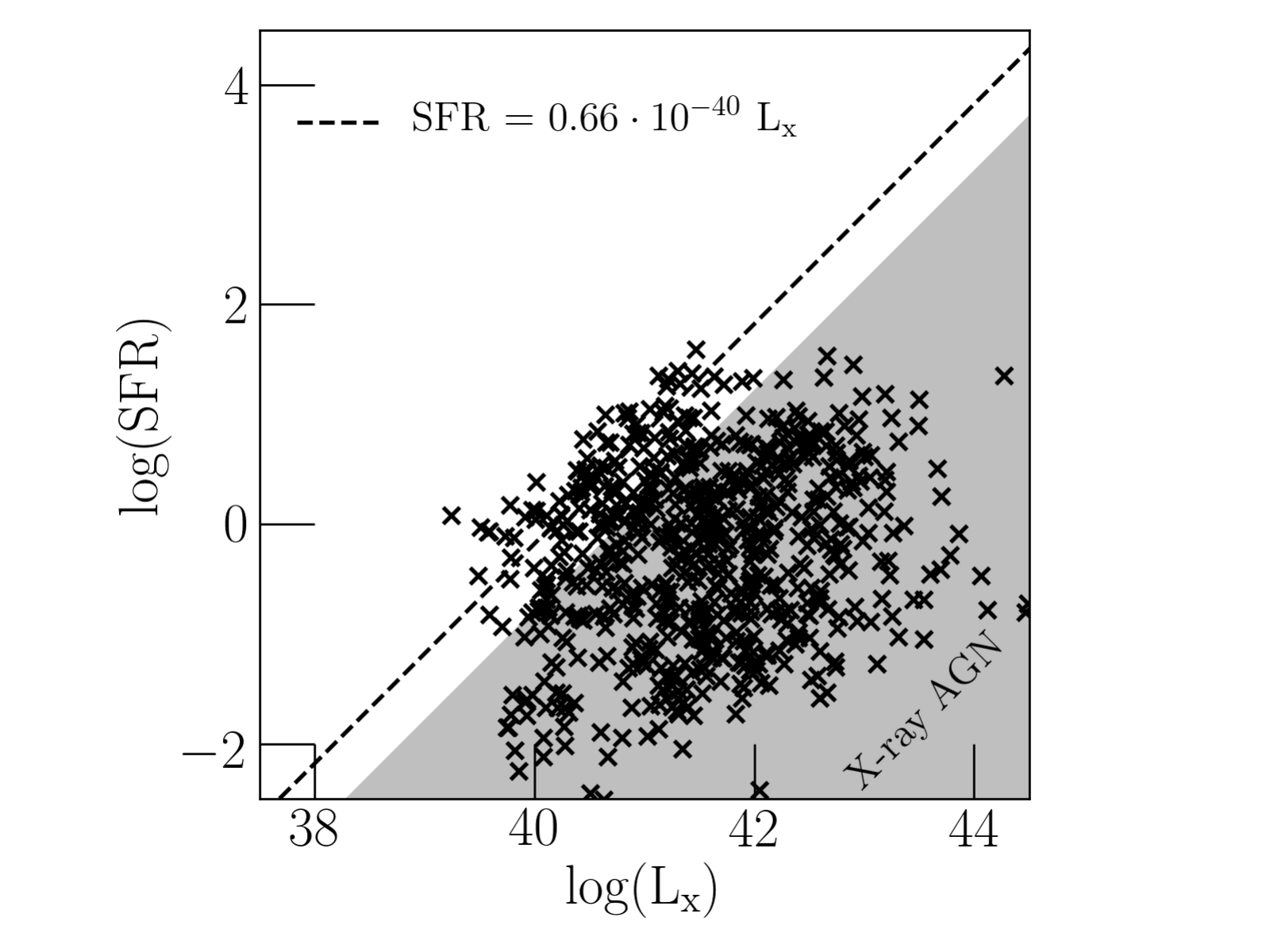}
                    \caption{Selection of X-ray AGN using the relation between star formation rate and X-ray luminosity. The dashed line shows the relation between SFR and $L_{\mathrm{X}}$ for non-AGN, derived in \citet{ranalli2003} and adapted here for use in the full band. The gray shaded region denotes the portion of the parameter space with excess X-ray emisssion due to the AGN. \label{fig:lxsfr}}
                    \end{figure}

 X-rays are produced by AGN via the Componization of UV/optical accretion disk photons by a corona of hot electrons \citep{haardt1991}. In galaxies, a variety of other sources produce X-ray emissions including X-ray binaries, and supernova remnants. In this work we select X-ray AGN by requiring X-ray emission to be present in excess of what is expected from non-AGN sources. Because massive stars are short-lived, the X-ray luminosity produced by their stellar end-products provides a measure of a galaxy's instantaneous star formation rate \citep{persic2007, lehmer2010, mineo2012}. 
 \citet{ranalli2003} determined the following empirical relations between star-formation rate ($M_{\odot}\ \mathrm{yr^{-1}}$) and X-ray luminosity (erg s$^{-1}$) in the soft (0.5--2 keV) and hard (2--10 keV) bands:  
    \begin{eqnarray}
    \mathrm{SFR}  &= 2.2\times 10^{-40}\ L_{0.5-2\ \mathrm{keV}} \\
    \mathrm{SFR} &= 2.0\times 10^{-40}\ L_{2-10\ \mathrm{keV}}
    \end{eqnarray}
 \citet{rg2009} found that soft X-rays more closely track recent SF episodes than hard X-rays, as measured via H$\alpha$ luminosity, but do not find such a discrepancy when comparing soft and hard X-rays with IR luminosities that trace SF activity averaged over the last 100 Myr, which is the same timescale used in computing SFRs in GSWLC-M1. Given that both soft and hard X-rays trace SF, we combine these two formulations to relate the star formation rates to the full-band (0.5--10 keV) X-ray luminosity, $L_{\mathrm{X}}$, by rearranging Equations 1 and 2 to
     \begin{eqnarray}
    L_{0.5-2\ \mathrm{keV}}   &= \mathrm{SFR}/2.2\times 10^{-40}  \\
    L_{2-10\ \mathrm{keV}}  &= \mathrm{SFR}/2.0\times 10^{-40} 
    \end{eqnarray}
 and summing them to get a relation between full-band luminosity and the SFR
 \begin{equation}
     L_{\mathrm{X}} = \mathrm{SFR}/2.2\times 10^{-40} +\mathrm{SFR}/2.0\times 10^{-40} 
 \end{equation}
which gives
\begin{equation} \label{eq:ranalli}
    \mathrm{SFR} = 1.05 \times 10^{-40}\ L_{\mathrm{X}} 
    \end{equation}
where SFRs are given for a Salpeter IMF. To convert to the Chabrier IMF \citep{chabrier2003} used in GSWLC, we divide SFRs by 1.58, which results in the following relationship:
    \begin{equation}
    \mathrm{SFR} = 0.66 \times 10^{-40}\ L_{\mathrm{X}}
    \end{equation}
 In Figure \ref{fig:lxsfr}, we show the adapted Ranalli relation and plot SFR versus $L_{\mathrm{X}}$ for the galaxies in our sample. Many galaxies lie on the relation, but even more have excess X-ray luminosities indicative of an AGN. \citet{ranalli2003} reported an intrinsic spread of 0.3 dex (1$\sigma$) for the original relation, which agrees with the scatter of our sample above the relationship. We take 0.6 dex (2$\sigma$) as the cutoff beyond which we conclude 549 are X-ray AGN (shaded area in Figure \ref{fig:lxsfr}).
 
 In previous studies with X-ray AGN in the star-forming region of the BPT diagram \citep{trouille2010, tbt2011, castello2012, pons2014, pons2016a}, X-ray AGN were selected by requiring $\log(L_{\mathrm{X}})>42$. \citet{pons2016b} use a less stringent requirement, $\log(L_{\mathrm{X}})>41$, based on a X-ray study of a subset of the 12 Micron Galaxy Survey \citep{brightman2011}, but do not consider SFRs in their selection of X-ray AGN, which likely leads to contamination. Our selection method is more comprehensive than the methods used in previous studies because it includes secure X-ray AGN at lower luminosities.  If $\log(L_{\mathrm{X}})>42$ were required for this study as in other studies, $\sim60\%$ of X-ray AGN would be eliminated.

    \section{Results}

    \subsection{Emission Line Diagnostics} \label{sec:bpt}
\begin{figure}[t!]
                    \epsscale{1.2}
                    \plotone{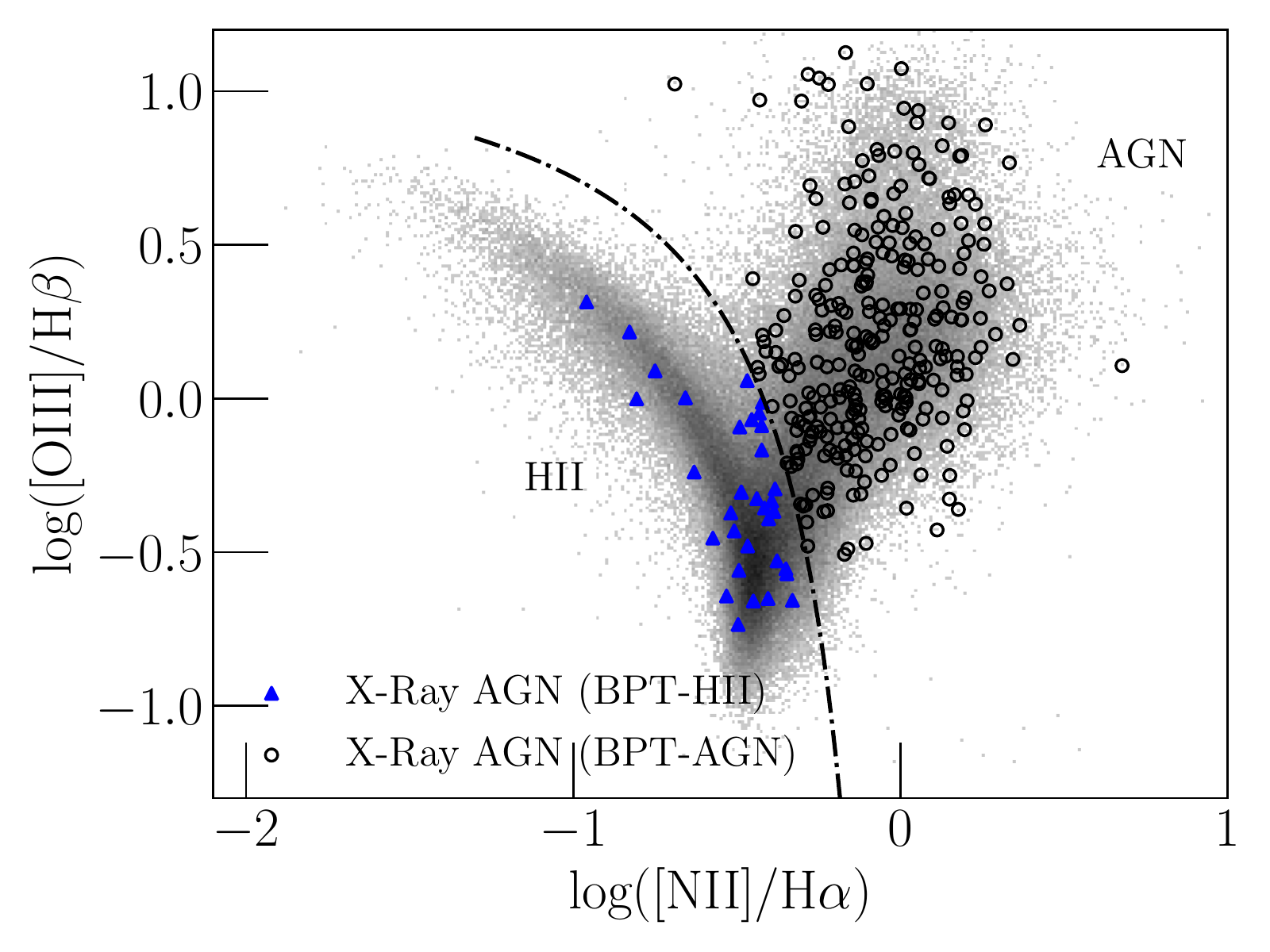}
                    \caption{Classical BPT Diagram showing [OIII]/H$\beta$ versus [NII]/H$\alpha$ with the empirical \citet{kauffmann2003} demarcation line. Of the 323 X-ray AGN detected in 3XMM, 34 (blue triangles) lie below the demarcation and 289 (black circles) above it. A two-dimensional histogram of galaxies from GSWLC-M1 make up the greyscale background, where the shading indicates the density. Unless otherwise noted, this same set of symbols and background galaxies are used in subsequent plots.\label{fig:mainbpt}}
                    \end{figure}

 Of 549 X-ray AGN, 323 have SNR$>$2 for all of the lines used in the BPT diagram (H$\alpha$, H$\beta$, [OIII], [NII]), while the remaining 226 (41\%) fail one or more SNR thresholds and cannot be reliably placed on the BPT diagram. SNR$>$2 for individual lines is sufficient for a reliable BPT classification \citep{juneau2014}. In Figure \ref{fig:mainbpt}, we show the BPT diagram, displaying 323 X-ray AGN. The \citet{kauffmann2003} (Ka03) demarcation line is shown and separates the star-forming (SF) and AGN branches. We find that 289 X-ray AGN are in the AGN region (black circles) and 34 X-ray AGN are found within the SF region (blue triangles). For this work, we refer to the first group as BPT-AGN and the second group as BPT-HII.

\begin{figure}[t!]
                    \epsscale{1.2}
                    \plotone{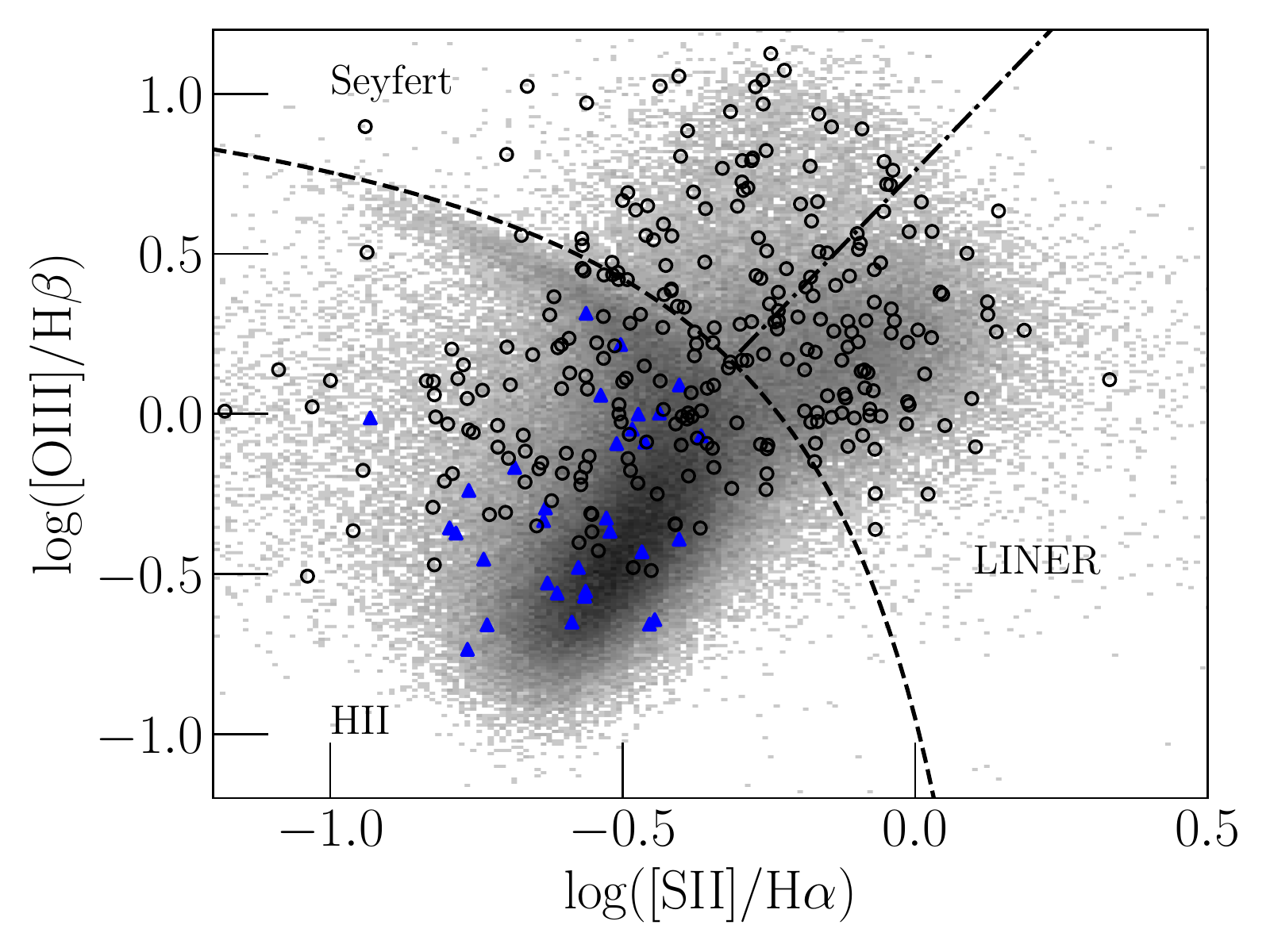}
                    \caption{[SII]/H$\alpha$ VO87 diagram, an alternate emission line diagnostic used for differentiating between LINERs, Seyferts, and Star-forming galaxies. This diagram mislabels even more BPT-AGN as star-forming galaxies. \label{fig:vo87-1}}
                    \end{figure}

\begin{figure}[t!]
                    \epsscale{1.2}
                    \plotone{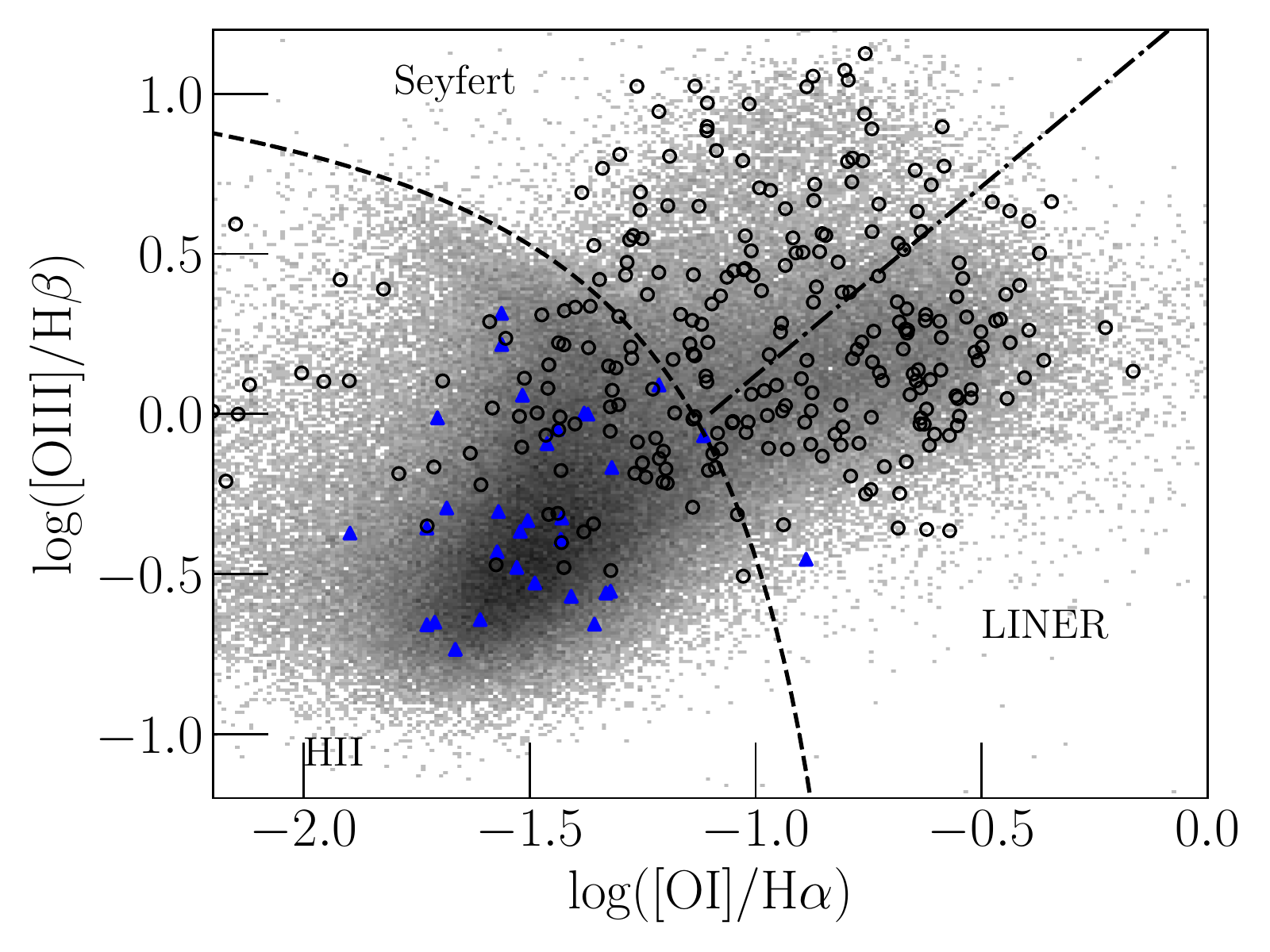}
                    \caption{[OI]/H$\alpha$ VO87 diagram. This diagnostic mislabels a similar number of more BPT-AGN as star-forming HII galaxies.\label{fig:vo87-2}}
                    \end{figure}

We also place the two groups of BPT-classified galaxies on two other commonly used optical emission-line diagnostics: the \citet{vo87} (VO87) diagrams, which keep [OIII]/H$\beta$ from the BPT diagram but swap [NII]/H$\alpha$ for [SII]/H$\alpha$ and [OI]/H$\alpha$. We wish to check if either of the VO87 diagrams can classify X-ray AGN more cleanly, i.e., with fewer of them in the star-forming (HII) region.  In Figures \ref{fig:vo87-1} and \ref{fig:vo87-2}, we show the VO87 diagrams for our two groups of X-ray AGN where [SII]/H$\alpha$ and [OI]/H$\alpha$ measurements were available, i.e., all lines had SNR$>2$. It is interesting to note that the BPT-HII remain classified by the VO87 diagrams as star-forming galaxies, suggesting that optical misclassification is not just an artifact of the regular BPT diagram. Among the BPT and VO87 diagrams, the BPT classifies the smallest portion of the X-ray AGN as star-forming galaxies, which corroborates its widely accepted use as a selection method for type 2 AGN. The VO87 diagrams are supposed to better distinguish LINERs than the BPT diagram \citep{kewley2006}. We see that X-ray AGN populate both the Seyfert and LINER regions and do so with similar density as the underlying sources. This suggests that both groups are powered by the same source, i.e., the AGN. We will return to the discussion of LINERs as AGN in Section \ref{sec:xrbpt}.

\subsection{Possible causes of misclassification}\label{sec:misclass}
 The present discrepancy in classification methods provides an opportunity to understand X-ray and optical properties of AGN. We present four principal reasons for this apparent misclassification, with some critical differences with respect to the scenarios presented in previous studies:

\begin{enumerate}
    \item {\it Star-formation dilution}. The spectroscopic fiber includes emission from star-forming regions that changes the resulting line ratios, turning what would otherwise be classified as a BPT-AGN galaxy into a BPT-HII. 
    \item {\it Broad-line contamination}. Classification is incorrect because the Balmer lines have a broad component, corrupting the line ratios and shifting the galaxy into the star-forming region of the BPT diagram.
    \item {\it AGN without emission lines}. X-ray AGN intrinsically has no line emission or the lines are heavily attenuated by dust, rendering these X-ray AGN unclassifiable by the BPT diagram were it not for the host SF. This scenario is related to the star formation dilution scenario, with the difference that the galaxy would not have been classified as an AGN even if there was no contamination from the host. One could argue that galaxies that would fall under this scenario are not really misclassified.
    \item {\it Low-ionization AGN}. In this scenario the lines indeed originate from an AGN and not from star formation, but the ionization levels are low enough that the galaxies fall below the \citet{kauffmann2003} AGN selection line. 
\end{enumerate}

 We explore the first two possibilities in separate sections, and the latter two in a joint section.

	\subsubsection{Dilution by star formation} \label{sec:ssfr}
\begin{figure}[t!]
                    \epsscale{1.2}
                    \plotone{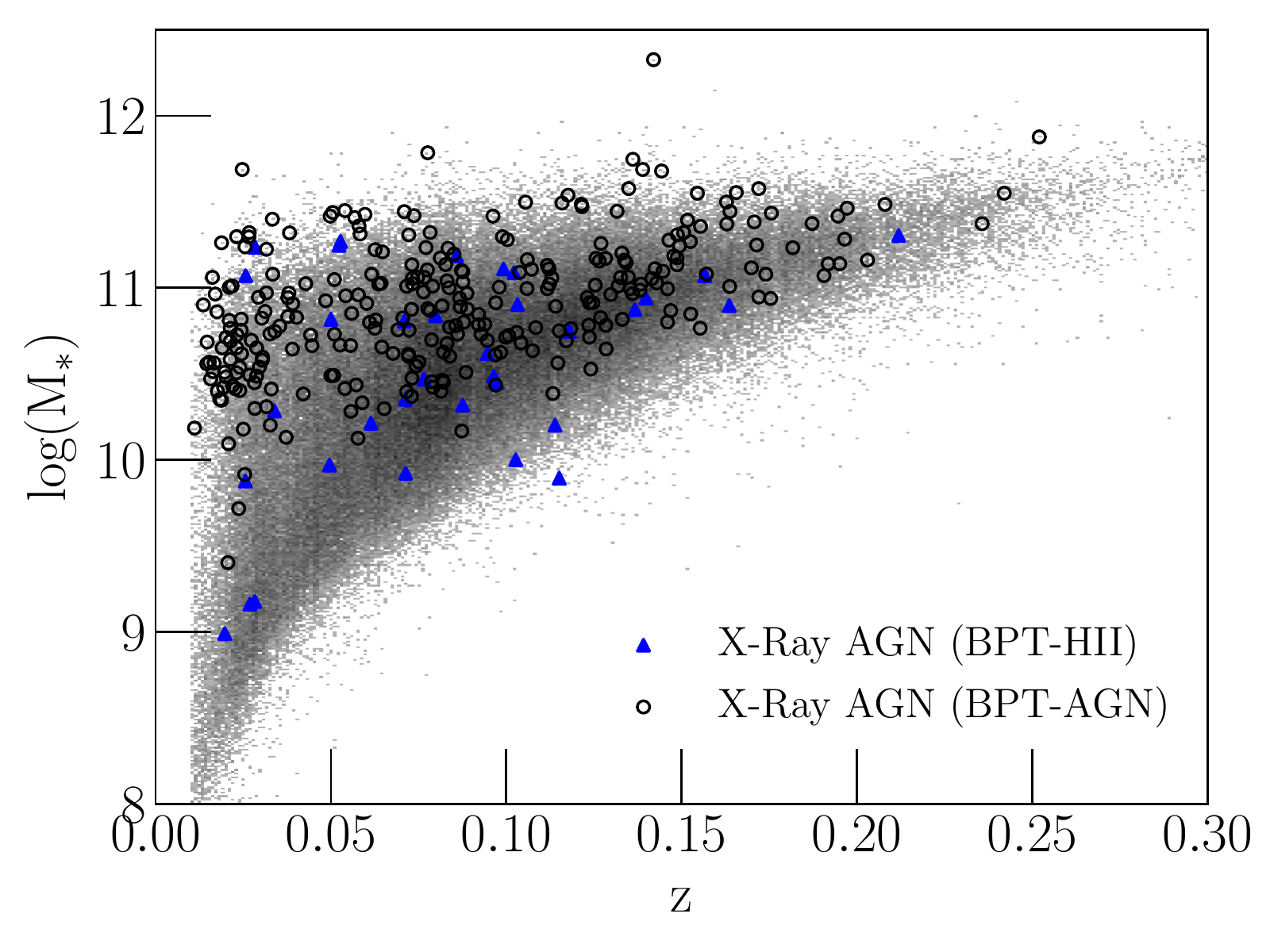}
                    \caption{Stellar mass of galaxies versus redshift. BPT-HII are also found in lower mass galaxies, but the correctly and ``incorrectly'' classified AGN overlap in mass. There is also no clear preference for redshift between the two groups of X-ray AGN. \label{fig:massz}}
                    \end{figure}

\begin{figure}[t!]
                    \epsscale{1.2}
                    \plotone{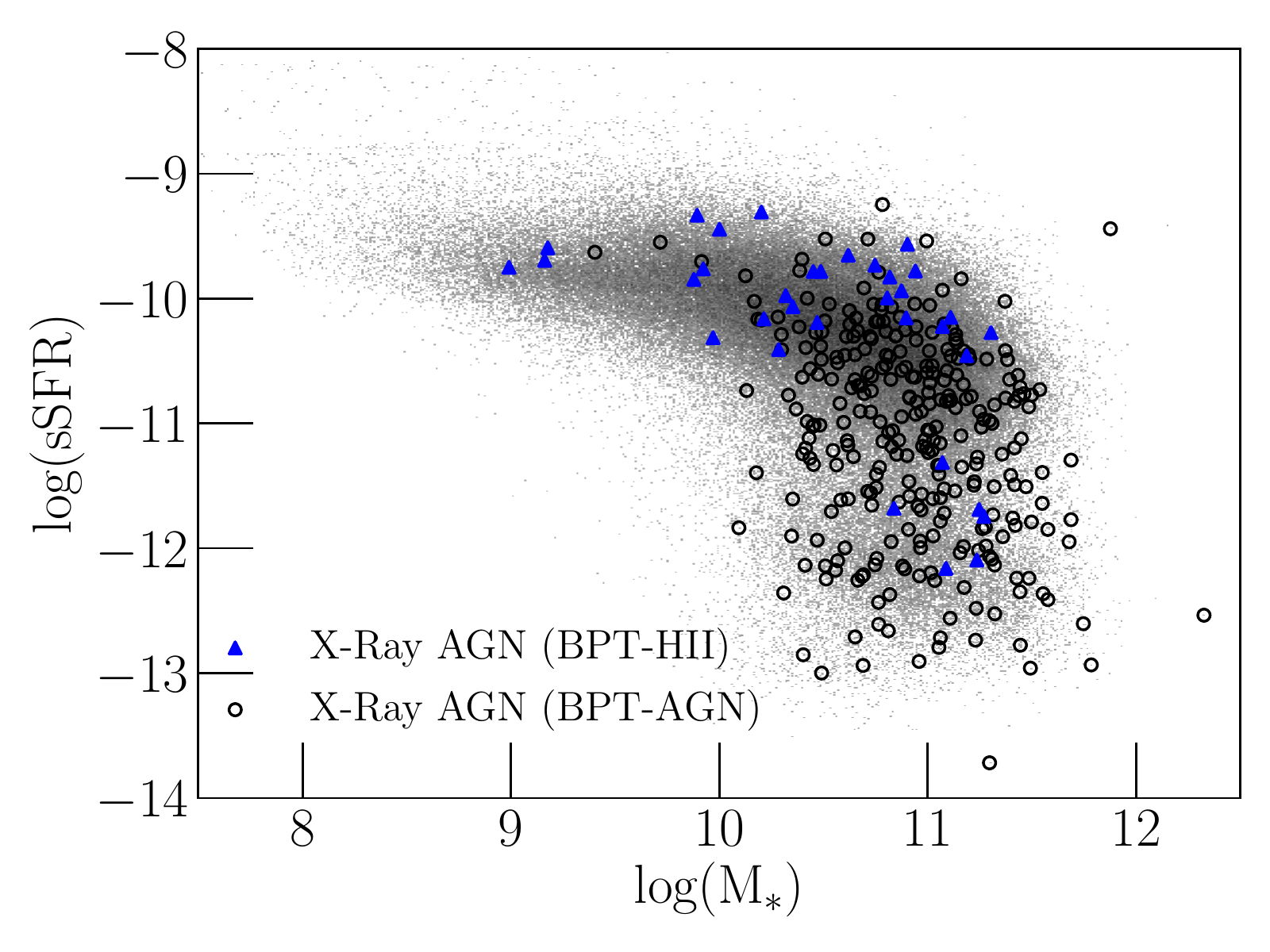}
                    \caption{Specific star formation rate versus stellar mass. The BPT-AGN span the entire range of sSFR while the BPT-HII are mostly concentrated in the blue portion of the galactic main sequence. However, there is no segregation by sSFR, making the star-formation dilution scenario an unlikely reason for ``misclassification''.  \label{fig:ssfrmass}}
                    \end{figure}

 One common explanation for the optical misclassification of X-ray AGN as BPT-HII is the dilution of nuclear emission by extra-nuclear light in a spectroscopic aperture. \citet{moran2002} used large spectroscopic apertures to demonstrate that many nearby type 2 AGN are diluted by star-formation from their hosts. The support for this scenario in our SDSS-based sample would come if one could demonstrate that some SFR-related or aperture-related quantity can separate correctly and incorrectly classified AGN. More distant galaxies will have a larger fraction of their light captured by the SDSS spectroscopic fiber, and would be more likely to be diluted. Similarly, smaller, or less massive galaxies would have a larger fraction of their total light in the fiber. We look at the BPT-HII and BPT-AGN in the mass-redshift plane in Figure \ref{fig:massz}. While BPT-HII have a tail of lower masses than BPT-AGN, the mass ranges overlap. Also, there is no separation in redshift, despite the detectability of nuclear activity depending on redshift \citep{moran2002}. Misclassified X-ray AGN are more likely to be found among lower-mass galaxies than their BPT-AGN counterparts but this difference may be intrinsic. In Figure \ref{fig:ssfrmass}, we address the question from another angle, by showing sSFR versus mass. Most of the BPT-HII are consistent with the upper blue portion of the main sequence whereas the BPT-AGN span the entire range of sSFRs, which reflects the typical sSFRs of the general population of BPT AGN, not just the X-ray AGN \citep{salim2007}. Furthermore, there appear to exist two populations of the BPT-HII where one smaller group has a much lower sSFR than the other, suggesting that star formation dilution cannot reliably explain their optical misclassification. The key point is that BPT-HII generally have higher sSFRs but so do many BPT-AGN and yet they are correctly classified. Figure \ref{fig:ssfrmass} shows sSFRs based on the total SFR. We have also looked at the SFR that we extrapolate to be present in the fiber, and the regions occupied by galaxies correctly and incorrectly classified still overlap.

\begin{figure}[t!]
                    \epsscale{1.25}
                    \plotone{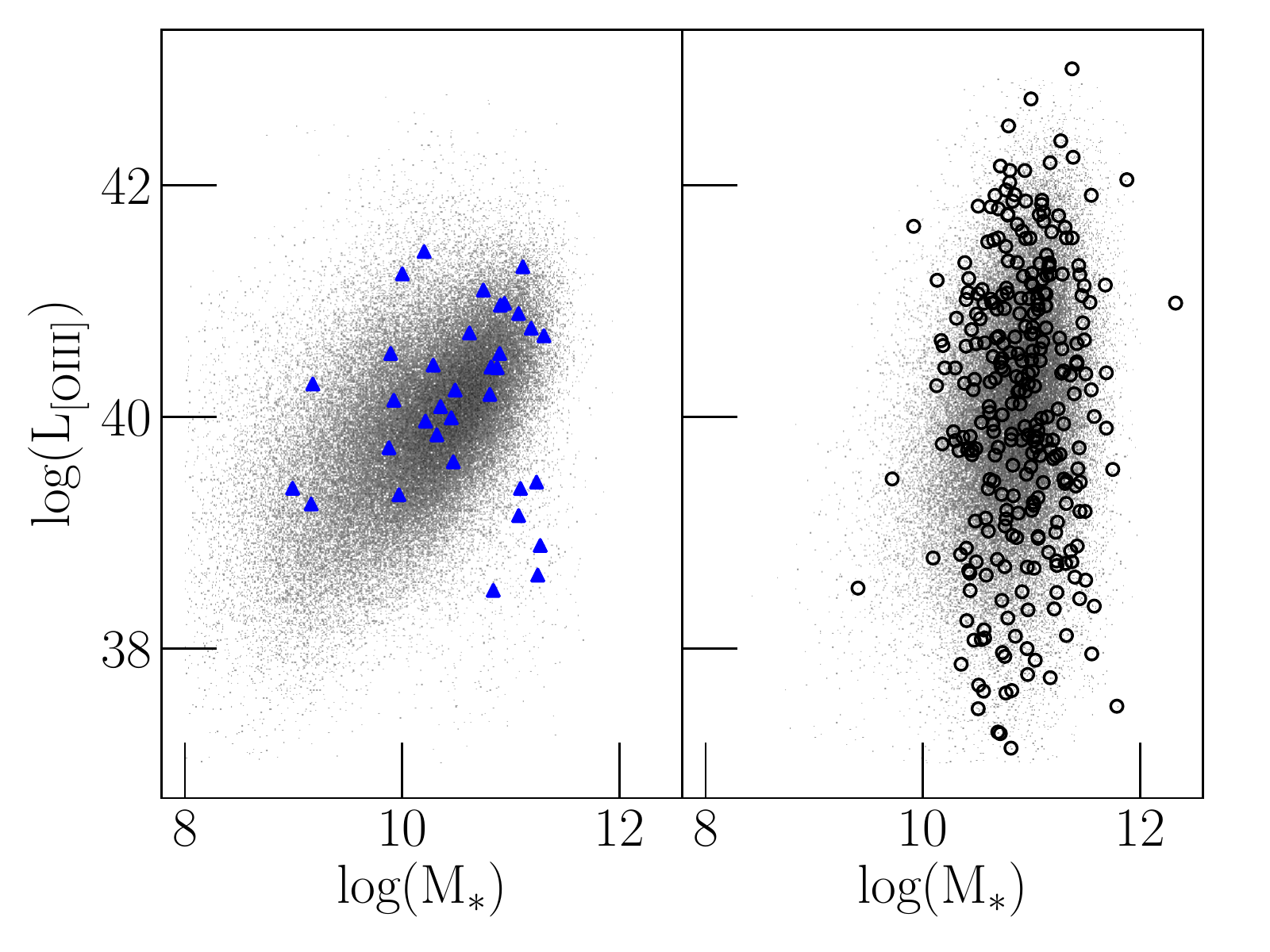}
                    \caption{De-reddened [OIII] luminosity versus stellar mass. (a) Left: BPT-HII are plotted on top of BPT star-forming galaxies from GSWLC-M1. (b) Right: BPT-AGN are plotted on top of BPT AGN from GSWLC-M1. $L_{\mathrm{[OIII]}}$ is an indicator of AGN strength. Six of the BPT-HII are not consistent with the distribution of the background star-forming galaxies but appear to be more consistent with the BPT AGN background. \label{fig:oiiilum}} 
                    \end{figure}

\begin{figure}[t!]
                    \epsscale{1.25}
                    \plotone{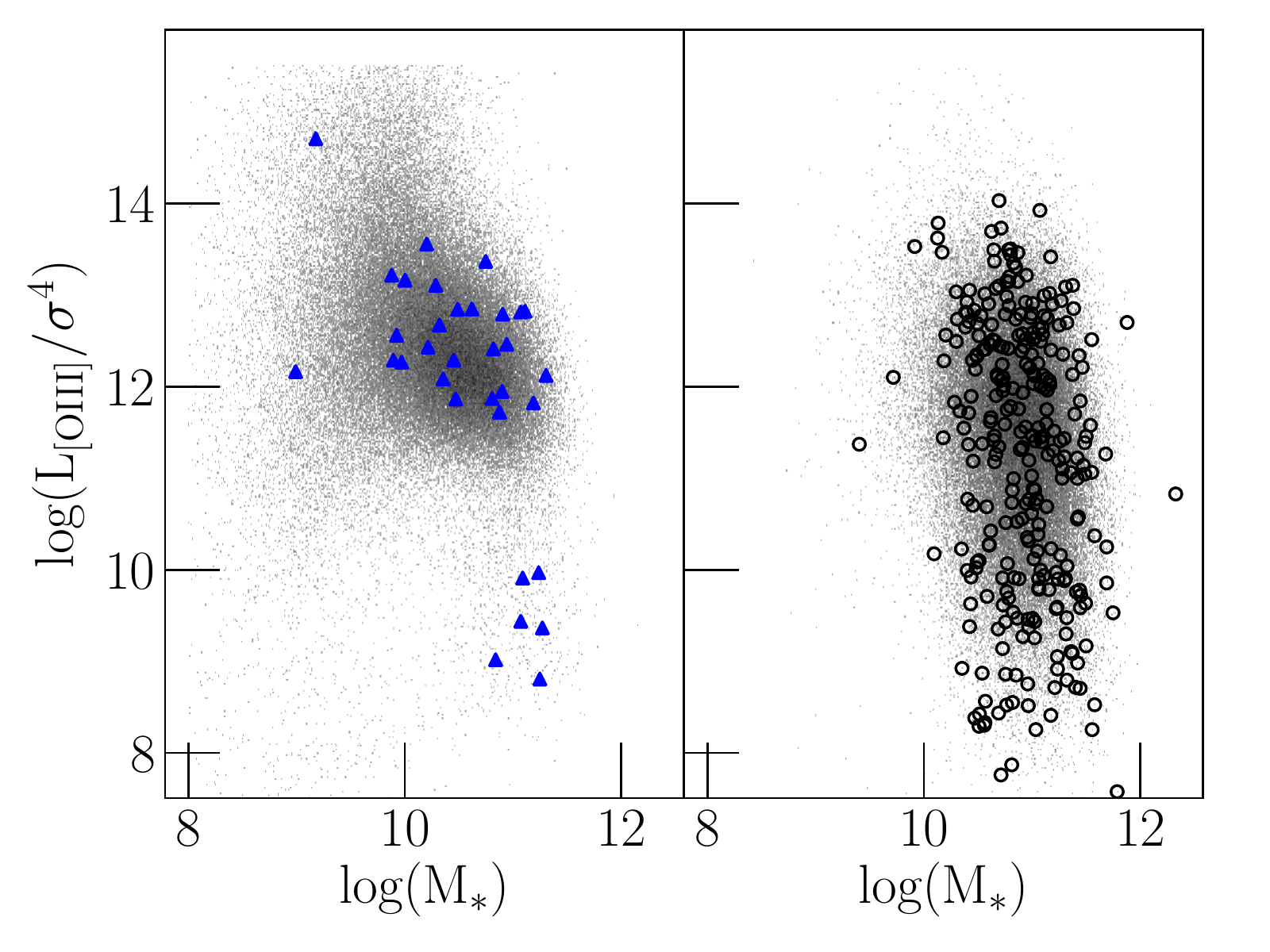}
                    \caption{Eddington Ratio Proxy ($L_{\mathrm{[OIII]}}$/$\sigma^4$) versus stellar mass.
                    (a) Left: BPT-HII are shown with the BPT star-forming galaxies from GSWLC-M1. (b) Right: BPT-AGN are plotted on top of the BPT AGN in GSWLC-M1. $L_{\mathrm{[OIII]}}$/$\sigma^4$ can provide a proxy for Eddington ratio of AGN if the contribution to the [OIII] line luminosity from star formation is minimal.\label{fig:eddrat}}
                    \end{figure}

  One factor that we did not take into account so far is that the strength of an AGN may affect how strong the corresponding emission from the narrow-line region is. Thus, it may be the case that lower luminosity or lower accretion rate AGN are more easily overwhelmed by star formation. Therefore, we next explore whether correctly and incorrectly classified X-ray AGN differ in their intrinsic AGN properties. One of the indicators of the accretion strength of an AGN in the optical regime is the luminosity of the forbidden [OIII] line,  $L_{\mathrm{[OIII]}}$, which scales with the bolometric luminosity of the AGN \citep{mulchaey1994, heckman2005, lamassa2010, azadimosdef2017, glikman2018}. A related quantity, $L_{\mathrm{[OIII]}}$ divided by the fourth power of the stellar velocity dispersion, $\sigma^4$, can provide an estimate of the Eddington ratio of the AGN \citep{kewley2006}. \citet{trump2015} suggested that higher sSFR galaxies must have more efficient AGN in order to be classified as BPT AGN because of the relative contributions from star formation. We show $L_{\mathrm{[OIII]}}$ and $L_{\mathrm{[OIII]}}$/$\sigma^4$ versus stellar mass in Figures \ref{fig:oiiilum} and \ref{fig:eddrat}, respectively, with the BPT star-forming galaxies from GSWLC-M1 composing the backgrounds in Figure \ref{fig:oiiilum}(a) and Figure \ref{fig:eddrat}(a) and the BPT AGN from GSWLC-M1 making up the backgrounds in Figures \ref{fig:oiiilum}(b) and \ref{fig:eddrat}(b). In Figures \ref{fig:oiiilum} and \ref{fig:eddrat}, there appear to be two distinct sequences for the two BPT types when mass is included, where in Figure \ref{fig:oiiilum}(a), the background BPT star-forming galaxies form one sequence where $L_{\mathrm{[OIII]}}$ appears to correlate with stellar mass, including galaxies with masses up to $\log(M_{*}) \approx 11$, while the star-forming galaxies in Figure \ref{fig:eddrat}(a) do not show a strong dependence of proxy Eddington ratio with stellar mass, which is expected because $L_{\mathrm{[OIII]}}$ from star-forming galaxies primarily comes from star-formation. The BPT AGN form sequences in Figures \ref{fig:oiiilum}(b) and \ref{fig:eddrat}(b) that have no apparent dependence on mass with masses ranging roughly from $10<\log(M_{*})<12$. The existence of these two sequences is apparently the case because of the different source of [OIII] emission that is dominant in BPT star-forming galaxies versus BPT AGN. The majority of the BPT-HII are more consistent with the background star-forming galaxies in Figures \ref{fig:oiiilum}(a) and \ref{fig:eddrat}(a) whereas the BPT-AGN all appear consistent with the background BPT AGN shown in Figures \ref{fig:oiiilum}(b) and \ref{fig:eddrat}(b), which is to be expected if the NLR dominates [OIII] emission. Six BPT-HII are more consistent with the background BPT AGN in Figures \ref{fig:oiiilum}(b) and \ref{fig:eddrat}(b) and, if the NLR is indeed the source of their [OIII] emission, their Eddington ratios are much lower than the average BPT-AGN, suggesting that they have relatively inefficient accretion.  We will return to these outliers in Section \ref{sec:nolines}.
  
   We find no considerable separation between the BPT-HII and BPT-AGN in terms of their $L_{\mathrm{[OIII]}}$, suggesting that, on its own, [OIII] emission is not a reliable tool for measuring nuclear activity of X-ray selected AGN because of possible contamination from star-formation. We find that $L_{\mathrm{[OIII]}}$/$\sigma^4$ provides a somewhat better separation between the BPT-AGN and BPT-HII than $L_{\mathrm{[OIII]}}$ but the improvement is marginal and the two groups still mostly overlap. Overall, we find $L_{\mathrm{[OIII]}}$ and $L_{\mathrm{[OIII]}}$/$\sigma^4$ are unreliable indicators of the strength of nuclear activity for X-ray selected AGN because of the overlapping ranges of the BPT star-forming and AGN sequences. We also confirm that at fixed SFR contained within the spectroscopic fiber, SFR$_{\mathrm{Fib}}$, none of the AGN-related quantities: $L_{\mathrm{X}}$, $L_{\mathrm{[OIII]}}$, or $L_{\mathrm{[OIII]}}/\sigma^{4}$ provides a separation between correctly and incorrectly classified objects, further disfavoring the SF dilution scenario.

   We conclude that SF dilution does not appear to be the most likely scenario for the majority (or even all) of the misclassified sources, given that there exist a large number of correctly classified BPT-AGN with similar sSFRs, mass, X-ray luminosity, and redshift.

    \subsubsection{Line Broadening}\label{sec:fwhm}
\begin{figure}[t!]
        \epsscale{1.3}
        \plotone{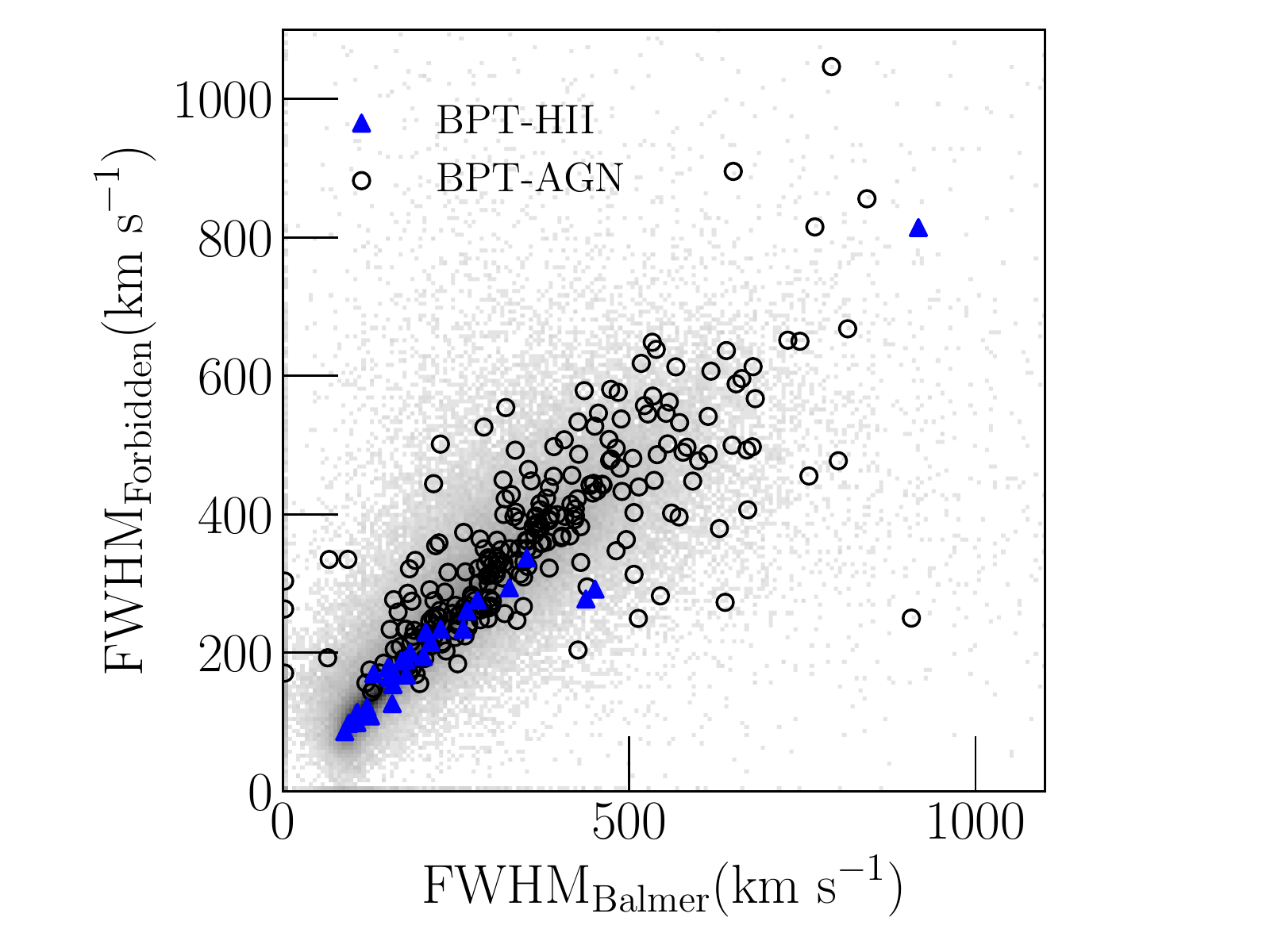}
        \caption{FWHM of forbidden lines versus FWHM of Balmer lines. The two FWHMs are usually comparable, so the Balmer line broadening does not seem like a viable explanation for misclassification of BPT-HII X-ray AGN. \label{fig:fwhm}}
    
        \end{figure}

 As discussed in Section \ref{sec:data}, type 1 AGN were excluded from the sample based on SDSS spectroscopic classification. Type 1 AGN have broadened Balmer lines and should therefore not be placed on emission line diagnostic diagrams, which are appropriate for the emission from NLRs. Furthermore, type 1 AGN are rare ($\sim$1\%) in general population of galaxies. However, a less extreme line broadening such that would not lead to spectroscopic classsification as a QSO may potentially affect the measured fluxes of Balmer series lines, including H$\alpha$ and H$\beta$, lowering the line ratios to make them appear as due to SF \citep{botte2004}. Instead of using some absolute line width threshold to identify any potential type 1 AGN remaining in our sample, a more relevant quantity in relation to the BPT diagram is to look at the comparison of the width of forbidden and of Balmer lines, following \citet{pons2014}. We show in Figure \ref{fig:fwhm} the FWHM of forbidden lines versus the FWHM of Balmer lines for the BPT-HII and BPT-AGN. FWHM values for Balmer lines were not available for 5 of the BPT-HII and 15 of the BPT-AGN. We manually inspected the SDSS DR7 spectra of these 5 BPT-HII and confirmed that these do not have any substantial broad lines. The S/N ratio of their Balmer lines is close to the cutoff and this may have prevented the measurements of FWHMs. Among the 29 BPT-HII in Figure \ref{fig:fwhm}, two show Balmer lines broadened by over 50\% compared to their forbidden lines. We have not tried to establish if this modest broadening is sufficient to affect their BPT classification. In any case, the majority of BPT-HII do not exhibit Balmer line broadening.

 It should be noted that emission line properties in the MPA/JHU catalog were derived using single Gaussian fits, which may underestimate the FWHM of certain lines with weak broad wings \citep{chavez2014}. We manually inspected the spectra of all 34 of our BPT-HII and did not find the Balmer lines in any of the spectra to have a substantial wing component, suggesting that our sample is not contaminated by type 1 AGN and the conclusions drawn in this section are valid.

\subsubsection{BPT-HII as AGN with intrinsically no emission lines or with very low ionization lines}\label{sec:nolines}

\begin{figure}[t!]
                    \epsscale{1.2}
                    \plotone{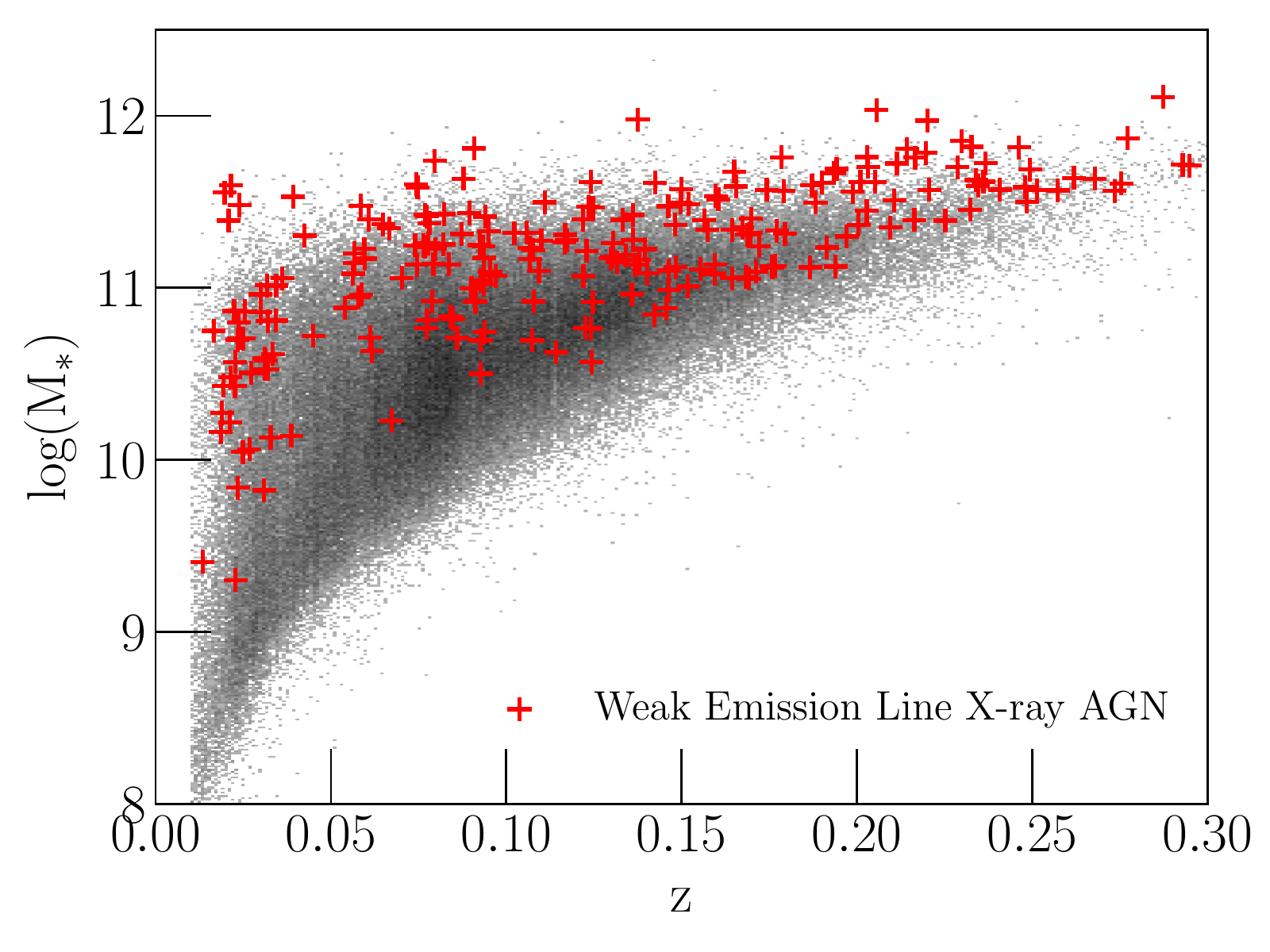}
                    \caption{Mass versus redshift for X-ray AGN that are unclassifiable by BPT diagram. Unclassifiable X-ray AGN are found in galaxies at similar distances and of similar mass to the BPT-AGN and BPT-HII, showing that they are not unclassifiable because they preferentially lie in low-mass galaxies or at large distances.
                    \label{fig:masszweakel}}
                    \end{figure}

\begin{figure}[t!]
                    \epsscale{1.2}
                    \plotone{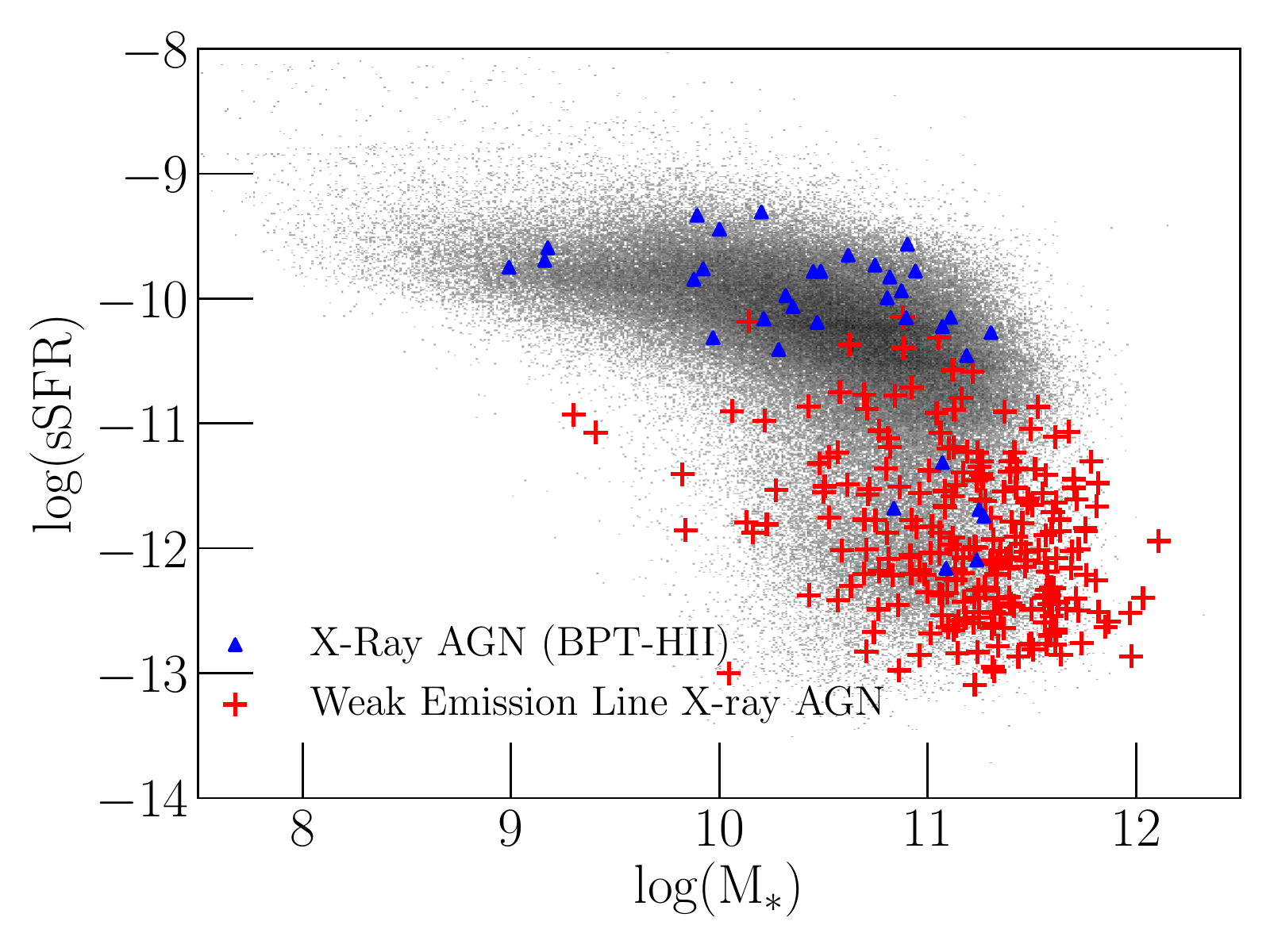}
                    \caption{sSFR versus stellar mass for BPT-HII and X-ray AGN that are unclassifiable by the BPT diagram. Unclassifiable X-ray AGN occupy the red, dead portion of the sSFR-Mass diagram. The BPT-HII and unclassifiable X-ray are mostly separated by their sSFRs. \label{fig:ssfrmassweakel}}
                    \end{figure}

We now turn our attention to a remarkable, but heretofore neglected fact that a large fraction of X-ray AGN, approximately 40\% in this study, are not classifiable using the BPT diagram, because one or more lines are not detectable. The reason for this may simply be that the spectroscopy is not of sufficient depth. In Figure \ref{fig:masszweakel}, we show that unclassifiable X-ray AGN span the same range of mass and redshift as either the BPT-HII or the BPT-AGN in Figure \ref{fig:massz}, demonstrating that they are not unclassifiable because they are more distant or are in less massive galaxies. The reasons for non-detection are thus intrinsic and may be a due to one or more of the following scenarios:
\begin{enumerate}
    \item AGN lack detectable NLRs because of obscuration by dust in the host galaxy \citep{rigby2006, goulding2009, trump2009, sargsyan2012}. 
    \item Given an AGN's measured X-ray luminosity, its expected optical emission is insufficient to be detected \citep{yan2011}.
    \item The AGN has a radiatively inefficient accretion flow (RIAF) which cannot adequately heat the NLR \citep{yuan2004, hopkins2009, trump2009}. 
    \item The NLR is not sufficiently heated because a complex NLR structure allows many ionizing photons to escape \citep{trouille2010}.
    \item The AGN has recently turned on and has not had enough time to heat the AGN such that it produces AGN lines \citep{schawinski2015}.
\end{enumerate}
If, however, such galaxies with undetectable AGN lines had sufficiently high star-formation, they will be (correctly) placed in the star-forming region of the BPT diagram. In Figure \ref{fig:ssfrmassweakel}, we show the sSFRs of the unclassifiable X-ray AGN and the ``misclassified'' ones and find that indeed the two groups are mostly distinct in terms of their sSFRs. We conclude that BPT-HII have AGN that are fundamentally similar to the AGN in unclassifiable X-ray AGN and that the intrinsically weak/non-detectable NLR is the most likely explanation for the majority of the so-called misclassified AGN. Considering that they would not be classifiable as AGN even if there was no SF, we argue it is not appropriate to refer to them as misclassified. 

For the smaller fraction of BPT-HII X-ray AGN, which have very low sSFRs and [OIII] properties more similar to low-ionization AGN in Figures \ref{fig:oiiilum}(b) and \ref{fig:eddrat}(b), the likely explanation is that the NLR emission of these AGN is intrinsically very similar to the emission of HII regions, leading to a genuine misclassification (our scenario 4 in Section \ref{sec:misclass}). These six BPT-HII may have low-ionization AGN incapable of heating the NLR enough for their nebular emission to be dominated by forbidden lines like BPT AGN or they may have NLRs heavily obscured by dust. A related possibility is that these six BPT-HII have AGN in the early stages of their duty cycle and have not yet had enough time to sufficiently heat and ionize their NLR.

\subsection{X-ray AGN fraction across the BPT Diagram}\label{sec:xrbpt}
        In the previous section, we concluded that most of the X-ray selected AGN found in the star-forming region of the BPT diagram are galaxies that would not be classifiable on the BPT diagram were it not for relatively high host sSFR. While having such AGN among the star-forming sample is not likely to affect emission line measurements (since the AGN contribution to the lines is insignificant), it does make the sample less pure, which may be of significance in studies that contrast galaxies with and without an AGN. In this section we aim to quantify the severity of this contamination, and do so by deriving the fraction of X-ray AGN across the BPT diagram. 
        
        Proper determination of the fraction of galaxies that are X-ray AGN needs to take into account that the {\it detectability} is a complex function of both the host/AGN properties and of the X-ray observations depth. We cannot simply take the X-ray fraction to be the raw fraction of X-ray AGN among all narrow-emission line galaxies in a given part of the BPT diagram, as this would dramatically underestimate the X-ray AGN fraction by including galaxies that could not have been detected in 3XMM, either because of host properties that differ from those of X-ray AGN hosts (e.g. lower mass) or because the candidate X-ray AGN is too distant to be detected in X-rays.

\begin{figure}[t!]
                    \epsscale{1.2}
                    \plotone{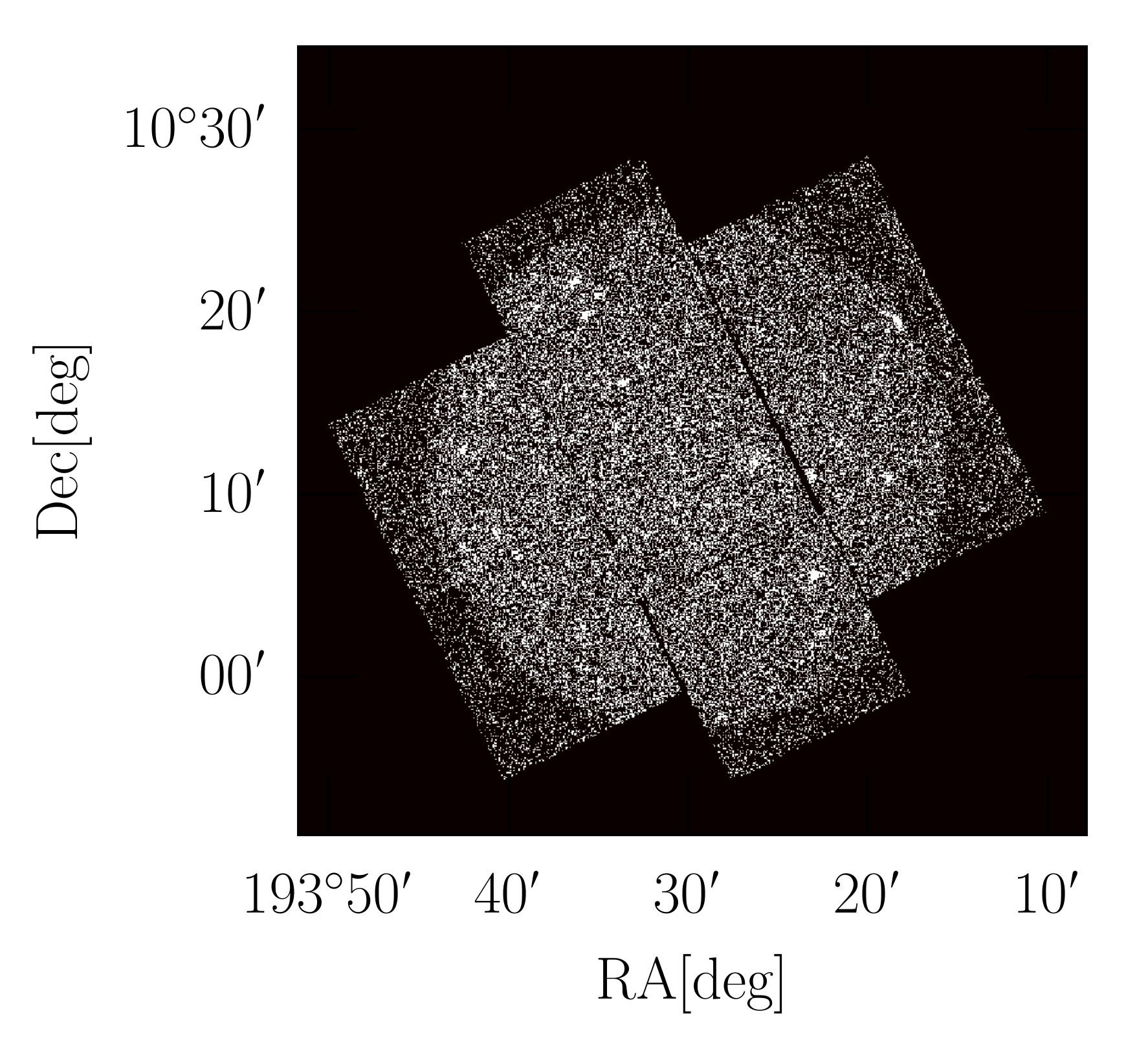}
                    \caption{One example field from 3XMM displaying binary image with pixels containing counts as `1's (white) and without counts as `0's (black). \label{fig:xrayobs}}
                    \end{figure}

        First, we need to identify which galaxies lie within the areas on the sky actually covered by 3XMM such that if these galaxies had significant X-ray emission, they would have been detected in 3XMM for the observation depth chosen for this study. From our X-ray sample, we take XMM observation IDs and download imaging data from the online XMM-Newton database for the M1, M2, and PN cameras, where available. We then use these X-ray images to create binary masks where any pixel with a count greater than 0 is set to 1. An example of a binary mask created from an observation field is shown in Figure \ref{fig:xrayobs}.

\begin{figure}[t!]
                    \epsscale{1.2}
                    \plotone{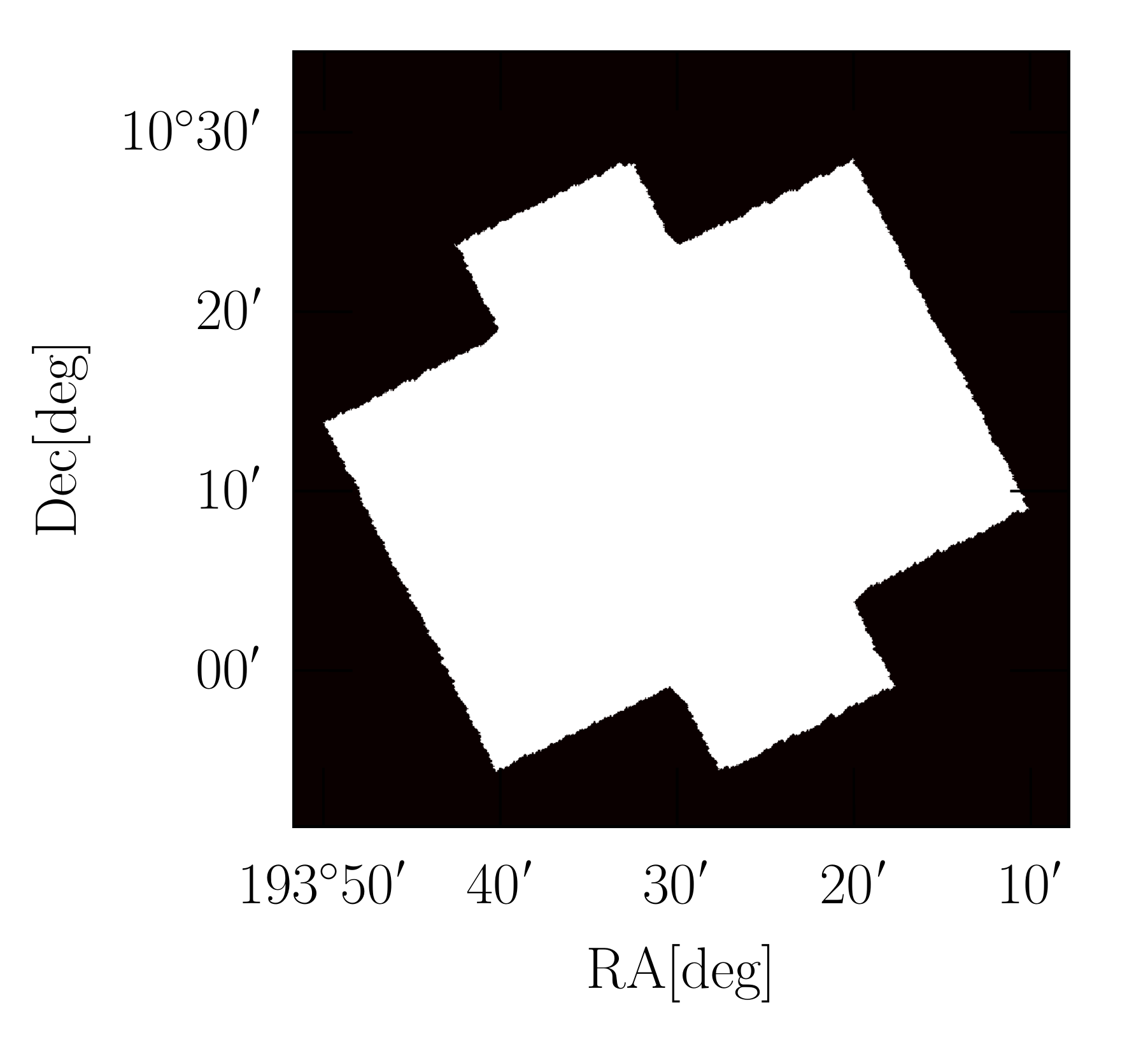}
                    \caption{Binary coverage map of field observed in Figure \ref{fig:xrayobs}. Pixels which had zero counts in the original observation have been filled in as white via a binary dilation which produces a cleaner, continuous field of observation for use in checking if the positions of GSWLC-M1 sources lie within these fields of view.
                    \label{fig:xraydil}}
                    \end{figure}

Within the actual field of view of the observations in Figure \ref{fig:xrayobs}, the number of pixels with 0 counts is quite substantial. To recover the actual area covered by X-ray observations, we used a binary dilation to fill in the pixels between counts and then used a binary erosion to remove the portions on the outside of the edges of the field (Figure \ref{fig:xraydil}). This resulted in binary maps of the portions of the sky which were covered by 3XMM and have exposure times in the range we selected for this study. 

\begin{figure*}[t!]
                    \epsscale{1.5}
                    \includegraphics[width=\linewidth]{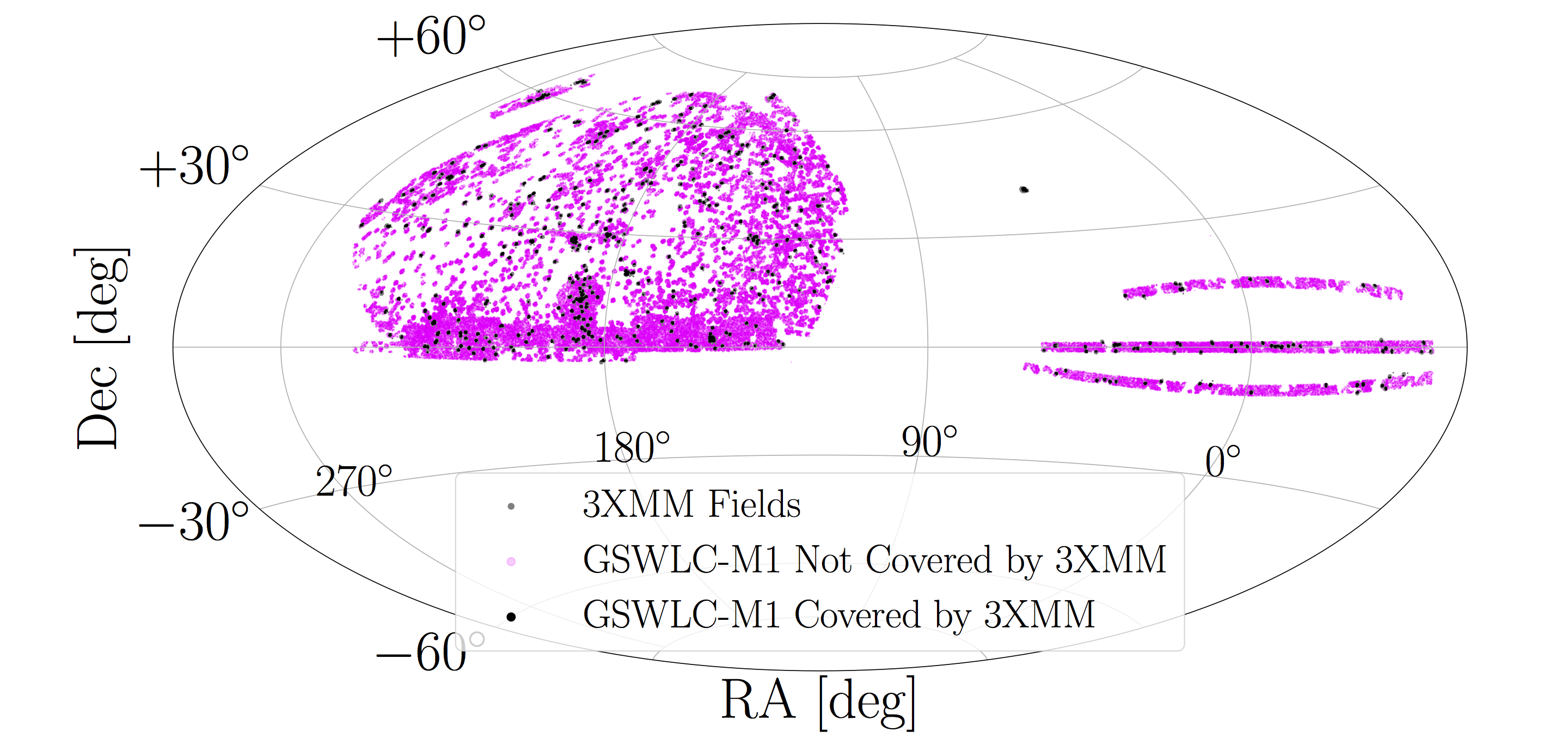}
                    \caption{Coverage map showing GSWLC-M1 galaxies and 3XMM fields. The 3XMM fields are shown in gray. Black stars are GSWLC-M1 galaxies that were included within the fields covered by 3XMM. Magenta galaxies are GSWLC-M1 galaxies not covered in 3XMM. \label{fig:xraycov}}
                    \end{figure*}

 We identify GSWLC-M1 galaxies covered by 3XMM by requiring the separation between the galaxy and any 3XMM pixel to be less than one pixel ($\sim1 \arcsec$). Only 6247 of the nearly 200,000 BPT-classifiable galaxies from GSWLC-M1 are found to be within the pointings of 3XMM and are shown on a map in Figure \ref{fig:xraycov}.

  To calculate the X-ray AGN fraction at different positions of the BPT diagram, we construct a distance metric ($d$) with which to measure the similarity of a galaxy with respect to a X-ray AGN, based on the stellar mass ($M_{*}$), redshift ($z$), log([NII]/H$\alpha$) ($x$), and log([OIII]/H$\beta$) ($y$), defined as 
    \begin{eqnarray}
    d = \Bigg[\left(\dfrac{x_{0}-x_{i}}{\sigma_{x}} \right)^2
     &+\left(\dfrac{y_{0}-y_{i}}{\sigma_{y}} \right)^2 + \nonumber\\ 
    \left(\dfrac{M_{*,0}-M_{*,i}}{\sigma_{M_{*}}} \right)^2  &+
    \left(\dfrac{z_{0}-z_{i}}{\sigma_{z}} \right)^2
    \Bigg]^{1/2} 
    \end{eqnarray}
 where each component is normalized by its respective standard deviation. We compute ``distances'' from each X-ray AGN ($x_{0}$, $y_{0}$, $M_{*,0}$, $z_{0}$) to each GSWLC-M galaxy ($x_{i}$, $y_{i}$, $M_{*,i}$, $z_{i}$) covered in 3XMM (including other X-ray AGN). Galaxies with components nearly matching an X-ray AGN's corresponding values will be considered similar if they lie within some maximum distance, $d_{\max}$, and the X-ray AGN fraction is the number of X-ray AGN divided by the number of all GSWLC-M galaxies within that distance. The latter are galaxies that have such host properties that they would have been detectable as X-ray AGN if they were one. We wish to use the minimum distance such that the fraction is reasonably well determined, and after some testing adopt a value of $d_{max}$ = 2.5.

\begin{figure}[t!]
                \epsscale{1.3}
                \plotone{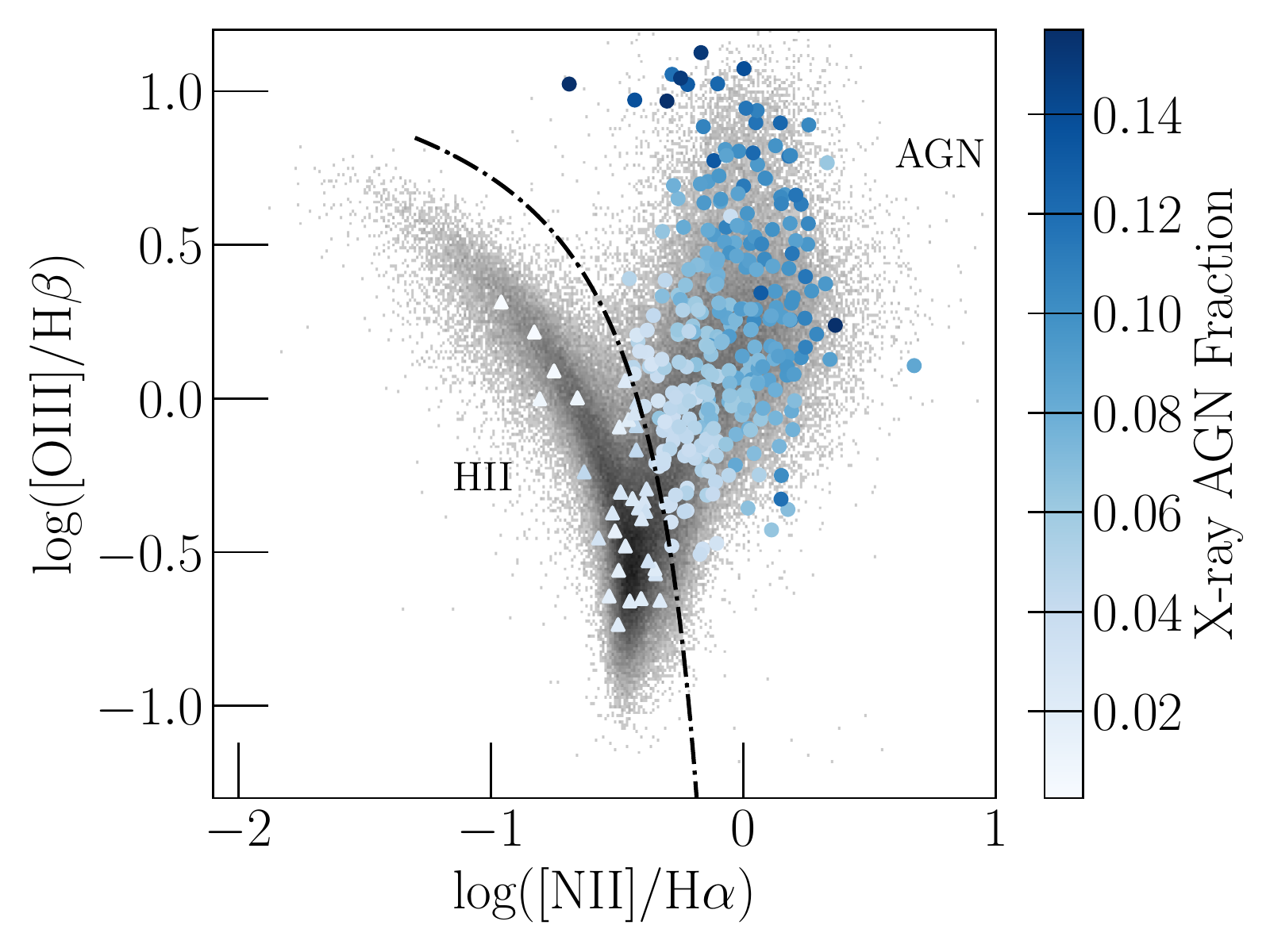}
                \caption{BPT diagram showing X-ray AGN color-coded by the X-ray AGN fraction among similar galaxies. The fraction of X-ray AGN in the star-forming sequence is quite low ($<2\%$).
                \label{fig:bptccode}}
                \end{figure}

 The above method allows us to probe the X-ray AGN fraction at the parameter space location of an X-ray AGN. The results are shown in Figure \ref{fig:bptccode} with the X-ray AGN color-coded by the X-ray fraction at their position in the BPT diagram. In the star-forming branch, the average X-ray AGN fractions for the upper ([OIII]/H$\beta >0$), middle ($-0.5<$[OIII]/H$\beta<0$), and lower ($-0.5<$[OIII]/H$\beta <0$) sections of the branch are 1\%, 2\%, and 3\%, respectively. In the AGN branch, the average X-ray AGN fractions for lower ($-0.5<$[OIII]/H$\beta<0$), middle ($0<$[OIII]/H$\beta<0.5$), upper ($0.5<$[OIII]/H$\beta<1$), and top ([OIII]/H$\beta>1$) are 5\%, 8\%, 10\%, and 14\%. Given our X-ray sample depth, these fractions pertain to AGN with log($L_{\mathrm{X}}) >$41. 
 
 Interestingly, the X-ray AGN fraction does not exceed 14\%, even at the uppermost tip of the AGN branch where all galaxies are genuine high-accretion AGN (Seyfert 2s) rather than LINERs, whose AGN nature has been debated. Furthermore, we find that the X-ray AGN fraction among LINERs is not anomalously low compared to the tip of the AGN branch, which is consistent with them being AGN.
 
We determine that the X-ray AGN contamination rate among the star-forming galaxies is only a couple of percent. However, considering that the X-ray AGN fraction does not exceed 14\% anywhere in the BPT diagram, it is possible that the actual contamination rate at the base of the SF branch is much higher if we consider 14\% to be the efficiency of finding X-ray AGN among BPT-AGN given our selected exposure time range for X-ray observations. In that case, the contamination rate in the base of the SF region may be larger by a factor of 100/14, which would correspond to an X-ray AGN fraction in the star-forming region of $\sim20\%$. Such contamination rate may present a problem for studies which require stringent removal of AGN.
 
    \section{Discussion} \label{sec:discussion}

  \citet{moran2002} showed that narrow-line AGN may be classified as BPT-star forming galaxies if the size of the spectroscopic aperture is large and is contaminated by star formation. This star-formation dilution hypothesis has played a prominent role in recent studies to explain the existence of some X-ray AGN which lack strong BPT AGN-like emission lines \citep{goulding2009, pons2014, pons2016a}. In Section \ref{sec:ssfr}, we find that the BPT-HII and BPT-AGN span similar mass and redshift ranges, suggesting that the classification is not strongly dependent on aperture effects and that star formation dilution cannot adequately explain misclassification, a result that is consistent with \citet{maragkoudakis2014} which found that [OIII]/H$\beta$ actually increased for some AGN when a larger aperture is used. Similarly, we do not find conclusive evidence in Section \ref{sec:ssfr} that a high sSFR alone will result in the misclassification of an X-ray AGN because there are nearly equal numbers of X-ray AGN with sSFR $>-10$ that are classified as BPT-AGN as there are BPT-HII. A high sSFR does, however, appear to be a requirement for most of the BPT-HII to be misclassifed.

 In this work, we attempted to use [OIII] luminosity to investigate the possibility that X-ray AGN lacking BPT-AGN line ratios have intrinsically weak or inefficient AGN, which may imply the presence of a RIAF. \citet{pons2014} and \citet{pons2016a} found low Eddington ratios, $\lambda <10^{-2}$, to be common among X-ray AGN which lack strong lines. At these lower Eddington ratios, a RIAF may be present but the AGN can still produce strong X-ray emissions through inverse Compton emission from the hot RIAF \citep{yuan2004}. In such a RIAF, the geometrically thin, optically thick Shakura-Sunyaev accretion disk is now absent or truncated and is replaced by a geometrically thick structure in which the inflow time is shorter than the radiative cooling time \citep{shakurasunyaev1973, narayan1994, narayan2005, ho2008}. In this work, we find six low sSFR BPT-HII with low values of $L_{\mathrm{[OIII]}}/\sigma^{4}$ compared to the general population of BPT AGN, suggesting low Eddington ratios and possibly the presence of a RIAF, if we assume their [OIII] emission comes primarily from the NLR. Because we suspect the majority of the BPT-HII do not have a detectable NLR, we cannot reliably use $L_{\mathrm{[OIII]}}/\sigma^{4}$ to comment on their accretion mode because of likely contamination from star-formation as we additionally find that BPT star-forming galaxies and AGN produce similar amounts of $L_{\mathrm{[OIII]}}$. This suggests that $L_{\mathrm{[OIII]}}$ may be an unreliable indicator of AGN strength for X-ray selected AGN, consistent with \citet{trouille2010} which found a dispersion of 2 dex in the the ratio of [OIII] to hard X-ray luminosities for narrow-line AGN. Recent works by \citet{jones2016} and \citet{thomas2018} likewise demonstrate the unreliability of $L_{\mathrm{[OIII]}}$, finding that contributions from SF can account for the majority ($\gtrsim90\%$) of emission from AGN lying just above the \citet{kauffmann2003} line as well as substantial portions ($10-20\%$) of the emission from the strongest AGN at the tip of the AGN branch, consistent with our finding that $L_{\mathrm{[OIII]}}$ spans the same range of values for BPT SF and AGN. It seems clear that optical indicators of AGN strength other than $L_{\mathrm{[OIII]}}$ like [NeIII] or similarly high ionization lines need to be statistically investigated in order to establish a reliable measure of an AGN's strength that can be employed in studies which utilize large samples of galaxies.

 In addition to star-formation dilution, it has been suggested that NLS1s could be mistaken as star-forming galaxies. In our sample we find no evidence of any NLS1. In contrast to the results in this paper, \citet{castello2012} find that most of their optically misclassified X-ray AGN are NLS1s with 600 km s$^{-1}<$FWHM(H$\beta$)$<$1200 km s$^{-1}$. To understand the cause of the differing results, we inspected the available SDSS spectra of the misclassified X-ray AGN from \citet{castello2012} and found that all but one of their galaxies are classified by the SDSS pipeline as `QSO's with most showing clearly broadened lines as well as significant blue continua and bright central sources. Therefore, the discrepancy in NLS1 fractions between \citet{castello2012} and ours is explained as a result of our difference in narrow-line galaxy selection: in that study, type 1 AGN were excluded solely by FWHM$_{\mathrm{H}\beta}=1200$ km s$^{-1}$ cut, whereas our work uses the SDSS spectroscopic classification to exclude type 1 AGN, which effectively eliminates galaxies with FWHMs broader than 500 km/s in any line. Similarly, \citet{pons2014}, which found 60\% of their optically misclassified X-ray AGN are NLS1, included such SDSS `QSO's in their sample. A more detailed study is required to determine precisely how broad-line components affect the position of an AGN in the BPT diagram. Our results suggest that using SDSS spectroscopic classification is a preferred way to exclude type 1 AGN than relatively high FWHM cuts.

  While understanding the causes of optical misclassification is indeed important, it is also pertinent to the study of AGN and of galaxy evolution as a whole to quantify the fraction of X-ray AGN at different positions in the BPT diagram. As we show in Section \ref{sec:xrbpt}, the X-ray AGN fraction in the star-forming branch of the BPT diagram is around 3\% at its base and decreases below 1\% towards the upper left side. This suggests that AGN contamination is not a severe issue. In addition to measuring the AGN contamination rate, we also find that the fraction of X-ray AGN is 14\% at the top of the AGN branch, where it is known that the galaxies must be true AGN. In contrast, \citet{azadimosdef2017} use very deep ($\sim100-4000$ ks) Chandra observations to find X-ray counterparts for $\sim$50\% of optically-selected AGN in the MOSDEF survey. In that work, the majority of the X-ray AGN have hard X-ray luminosities in the range 10$^{43}$-10$^{45}$ erg s$^{-1}$ and L$_{\mathrm{[OIII]}}>10^{42}$ erg s$^{-1}$, whereas the majority of our X-ray AGN have full-band luminosities below $10^{43}$ erg s$^{-1}$ and L$_{\mathrm{[OIII]}}<10^{42}$ erg s$^{-1}$, and it seems reasonable that a higher recovery fraction would be found for optically selected AGN of higher luminosities. Our low recovery fraction of $14\%$ suggests that, despite X-ray selection being able to provide secure AGN, there are populations of AGN that this method misses because of the complex structure of gas and dust surrounding an AGN that prevents significant X-ray emission from escaping, as is the case for AGN surrounded by Compton-thick clouds in galaxy mergers \citep{ricci2017, satyapal2017}. The number of such Compton-thick AGN found in galaxy mergers increases at higher redshifts \citep{kocevski2015, lanzuisi2015, delmoro2016, koss2016}, where X-ray selection of AGN is often the only method available as optical emission lines get redshifted further into the infrared. Our results suggest that it is imperative that other methods for selection of AGN at higher redshifts be employed or developed to address this problem. The TBT diagram provides a selection method for doing so optically up to $z\sim1.4$ but the lines required for using it are fairly weak and may be difficult to detect. Fortunately, recent works utilizing the MOSDEF survey have extended the utilization of BPT diagrams to higher redshifts \citep{coilmosdef2015, kriekmosdef2015} and have investigated the reliability of such diagnostics at these redshifts \citep{azadimosdef2017}.

 \section{Conclusions}
In this work, we used 323 X-ray AGN in relatively massive hosts, at redshifts below 0.3, and spanning a wide range of X-ray luminosities to investigate their classification in the BPT diagram and to quantify the X-ray AGN fraction as a function of position in the BPT diagram. We summarize the main results.
\begin{enumerate}
    \item A fraction of X-ray AGN is found in the star-forming region of the BPT diagram, as shown in other studies. 
    \item At fixed specific SFR, none of the AGN-related quantities of $L_{\mathrm{X}}$, $L_{\mathrm{[OIII]}}$, and $L_{\mathrm{[OIII]}}/\sigma^4$ provides a separation between AGN classified as such by the BPT diagram and the ones apparently misclassified as star-forming galaxies, which suggests that the star-formation dilution scenario is not consistent with observations.
    \item The use of $L_{\mathrm{[OIII]}}$ as an indicator of nuclear activity of AGN is not reliable because $L_{\mathrm{[OIII]}}$ from BPT star-forming galaxies overlaps significantly with that of BPT AGN.
    \item Furthermore, nearly 40\% of X-ray AGN, having a full range of X-ray luminosities, are not detectable in one or more of BPT lines. This also speaks to the limitations of $L_{\mathrm{[OIII]}}$ as an indicator of AGN strength.
    \item We conclude that the majority of X-ray AGN classified by BPT as star-forming do not have a narrow-line region and are only classifiable in the BPT diagram because of the contributions from host star-formation. Thus, the intrinsic properties of the X-ray AGN in the star-forming region of the BPT diagram are closer to X-ray AGN that have no emission lines (result 4) than the ones on the AGN branch.
    \item  The X-ray AGN fraction is 1\% at the tip of the star-forming branch and 3\% at its base. Selection of star-forming galaxies by the BPT diagram provides a reasonably clean sample of galaxies, mostly unaffected by X-ray AGN.
    \item The X-ray AGN fraction at the tip of the AGN branch is only 14\%, which suggests that X-ray selection of BPT AGN is largely incomplete even with $\sim$20 ks exposure times, which are deep for our low redshifts.
    
    \item The X-ray AGN fraction is similar in LINER and Seyfert regions of the BPT diagram. Furthermore they are found equally among the Seyfert and LINER populations in the VO87 emission-line diagrams. These results are consistent with LINERs being AGN. 
    
\end{enumerate}
\section{Acknowledgements} \label{sec:acknowl}
This research has made use of data obtained from the 3XMM XMM-Newton serendipitous source catalogue compiled by the 10 institutes of the XMM-Newton Survey Science Centre selected by ESA. This research made use of Astropy, a community-developed core Python package for Astronomy \citep{astropy2013}.

The authors would like to thank Robert E. Butler for his thoughtful questions and helpful comments which improved the analysis in this work. Additionally, we thank Roberto Terlevich, Mackenzie L. Jones, and Mojegan Azadi for bringing to our attention literature which improved the quality of the discussion of our results in the context of previous works, as well as the anonymous referee whose comments helped us improve the clarity and focus of this work.
\bibliographystyle{aasjournal}
\bibliography{refs.bbl}

\end{document}